\documentclass[aps,twocolumn,groupedaddress,a4paper]{revtex4}
\pdfoutput=1

\newcommand\T{\rule{0pt}{2.6ex}}
\newcommand\B{\rule[-1.2ex]{0pt}{0pt}}

\usepackage {latexsym, graphicx,array, subfigure,amsmath,setspace,color,amssymb,amsfonts}

\begin{document}

\title{Directional States of Symmetric-Top Molecules Produced by Combined Static and Radiative Electric Fields}
\author{Marko H\"{a}rtelt and Bretislav Friedrich}
\affiliation{Fritz-Haber-Institut der Max-Planck-Gesellschaft, Berlin, Germany}
\date{\today}

\begin{abstract}
We show that combined electrostatic and radiative fields can greatly amplify the directional properties, such as axis orientation and alignment, of symmetric top molecules. In our computational study, we consider all four symmetry combinations of the prolate and oblate inertia and polarizability tensors, as well as the collinear and perpendicular (or tilted) geometries of the two fields. In, respectively, the collinear or perpendicular fields, the oblate or prolate polarizability interaction due to the radiative field forces the permanent dipole into alignment with the static field. Two mechanisms are found to be responsible for the amplification of the molecules' orientation, which ensues once the static field is turned on: (a) permanent-dipole coupling of the opposite-parity tunneling doublets created by the oblate polarizability interaction in collinear static and radiative fields; (b) hybridization of the opposite parity states via the polarizability interaction and their coupling by the permanent dipole interaction to the collinear or perpendicular static field. In perpendicular fields, the oblate polarizability interaction, along with the loss of cylindrical symmetry, is found to preclude the wrong-way orientation, causing all states to become high-field seeking with respect to the static field. The adiabatic labels of the states in the tilted fields depend on the adiabatic path taken through the parameter space comprised of the permanent and induced-dipole interaction parameters and the tilt angle between the two field vectors.
\end{abstract}

\maketitle

\section{Introduction\label{interactions}}

Directional states of molecules are at the core of all methods to manipulate molecular trajectories. This is because only in directional states are the body-fixed multipole moments ``available'' in the laboratory frame where they can be acted upon by space-fixed fields.

In the case of polar molecules, the body-fixed permanent dipole moment is put to such a full use in the laboratory by creating oriented states characterized by as complete a projection of the body-fixed dipole moment on the space-fixed axis as the uncertainty principle allows.

The classic technique of \emph{hexapole focusing} selects precessing states of symmetric top molecules (or equivalent) which are inherently oriented by the first-order Stark effect \cite{special}, independent of the strength of the electric field applied (at low fields). For the lowest precessing state, only a half of the body-fixed dipole moment can be projected on the space fixed axis and utilized to manipulate the molecule. \emph{Pendular orientation} is more versatile \cite{Friedrich1991,LOESCH1990}, applicable to linear molecules as well as to symmetric and asymmetric tops, but works well only for molecules with a large value of the ratio of the body-fixed electric dipole moment to the rotational constant. However, for favorable values of this ratio, on the order of $100$ Debye/cm$ ^{-1}$, the ground pendular state of a $^{1}\Sigma $ molecules utilizes up to $90$\% percent of the body-fixed dipole moment in the space-fixed frame at feasable electric field strength, of the order of $100$ kV/cm.

Pendular states of a different kind can be produced by the induced-dipole interaction of a nonresonant laser field with the anisotropic molecular polarizability \cite{FRIEDRICH1995,FRIEDRICH1995a,Colloquium2003,LasConManMol}. The resulting directional states exhibit alignment rather than orientation, and can be used to extend molecular spectroscopy \cite{Friedrich1996}, suppress rotational tumbling \cite{Kim1997,Larsen2000,Poulsen2004}, focus molecules \cite{Zhao2003}, or to decelerate and trap molecules \cite{Fulton2006}.

The reliance on special properties of particular molecules has been done away with by the development of techniques that \emph{combine a static electric field with a nonresonant radiative field}. The combined fields give rise to an amplification effect which occurs for any polar molecule, as only an anisotropic polarizability, along with a permanent dipole moment, is required. This is always available in polar molecules. Thus, for a number of molecules in their rotational ground state a very weak static electric field can convert second-order alignment by a laser into a strong first-order orientation that projects about $90$\% of the body-fixed dipole moment on the static field direction. If the polar molecule is also paramagnetic, combined static electric and magnetic fields yield similar amplification effects. So far, the combined-fields effects have been worked out for linear polar molecules, and corroborated in a number of experiments by the Buck \cite{Baumfalk2001,Buck03-01,Friedrich2003} and Farnik groups \cite{farnik} on linear rare-gas clusters, as well as by the group of Sakai who undertook a detailed femtosecond-photoionization study on oriented OCS \cite{Sakai2003,Tanji2005}.

Here we analyze the amplification effects that combined electrostatic and nonresonant radiative fields bring out in \emph{symmetric top} molecules. In our analysis we draw on our previous work on linear molecules \cite{FriedHerschb99-1,FriedHerschb99-2} as well as on the work of Kim and Felker \cite{KimFelker98-1}, who have treated symmetric top molecules in pure nonresonant radiative fields.

For linear molecules, the amplification of the orientation arises from the coupling by the electrostatic field of the members of quasi-degenerate opposite-parity tunneling doublets created by the laser field. Thereby, the permanent electric dipole interaction creates oriented states of indefinite parity \cite{Friedrich2003}. We find that this mechanism is also in place for a class of precessing symmetric top states in the combined fields. However, another class of precessing symmetric top states is prone to undergo a sharp orientation via a different mechanism: the laser field alone couples states with opposite parity, thereby creating indefinite parity states that can then be coupled by a weak electrostatic field to the field's direction.

In Section \ref{sec:SymTopInt}, we set the stage by writing down the Hamiltonian (in reduced, dimensionless form) for a symmetric top molecule in the combined electrostatic and linearly polarized nonresonant radiative fields, and consider all four symmetry combinations of the polarizability and inertia tensors as well as the collinear and perpendicular (tilted) geometries of the two fields. We derive the elements of the Hamiltonian matrix in the symmetric top basis set and discuss their symmetry properties. We limit ourselves to the regime when the radiative field stays on long enough for the system to develop adiabatically. This makes it possible to introduce adiabatic labelling of the states in the combined fields, sort them out systematically, and define their directional characteristics, such as orientation and alignment cosines.

In Section \ref{sec:EffPot} we introduce an effective potential that makes it possible to regard the molecular dynamics in the combined fields in terms of a 1-D motion. The effective potential is an invaluable tool for making sense of some of the computational results obtained by solving the eigenproblem in question numerically.

In Section \ref{sec:Disc} we present the results proper. First, we construct correlation diagrams between the field free states of a symmetric top and the harmonic librator, which obtains at high fields. Then we turn to the combined fields, and consider collinear fields and perpendicular fields in turn. It is where we discuss the details of the two major mechanisms responsible for the amplification of the orientation by the combined fields.

Section \ref{sec:Disc} discusses the possibilities of applying the combined fields to a swatch of molecules (representing the symmetry combinations of the polarizability and inertia tensors) and of making use of the orientation achieved in a selection of applications. The main conclusions of the present work are summarized in Section \ref{sec:Concl}.

\section{Symmetric-top molecules and their interactions with static and radiative electric fields\label{sec:SymTopInt}}

We consider a symmetric-top molecule which is both \emph{polar} and \emph{polarizable}. The inertia tensor, $\mathbf{I}$, of a symmetric top molecule possesses a three-fold or higher axis of rotation symmetry (the figure axis), and is said to be prolate or oblate, depending on whether the principal moment of inertia about the figure axis is smaller or larger than the remaining two principal moments (which are equal to one another). The principal axes $a,b,c$ of $\mathbf{I}$ are defined such that the principal moments of inertia increase in the order $I_{a}<I_{b}<I_{c}$.

The symmetry of the inertia tensor is reflected in the symmetry of the polarizability tensor, $\boldsymbol{\alpha}$, in that the principal axes of the two tensors are collinear. However, a prolate or oblate inertia tensor does not necessarily imply a prolate or oblate polarizability tensor. Four combinations can be distinguished: (i) $\mathbf{I}$ prolate and $\boldsymbol{\alpha }$ prolate, i.e., $I_{a}<I_{b}=I_{c}$ and $\alpha _{a}<\alpha _{b}=\alpha _{c}$ with $a$ the figure axis; (ii) $\mathbf{I}$ prolate and $ \boldsymbol{\alpha }$ oblate, i.e., $I_{a}<I_{b}=I_{c}$ and $\alpha _{a}>\alpha _{b}=\alpha_{c}$, with $a$ the figure axis; (iii) $\mathbf{I}$ oblate and $ \boldsymbol{\alpha }$\ prolate, i.e., $I_{a}=I_{b}<I_{c}$ and $\alpha _{a}=\alpha _{b}>\alpha _{c}$ with $c$ the figure axis; (iv) $\mathbf{I}$ oblate and $\boldsymbol{\alpha }$ oblate, i.e., $I_{a}=I_{b}<I_{c}$ and $\alpha_{a}=\alpha _{b}<\alpha _{c}$, with $c$ the figure axis. The body-fixed permanent electric dipole moment $\boldsymbol{\mu} $ of a symmetric-top molecule is bound to lie along the molecule's figure axis. The symmetry combinations of $ \mathbf{I}$ and $\boldsymbol{\alpha}$\ are summarized in Table \ref{tbl:SymComb} in terms of the
rotational constants 
\begin{equation}
A\equiv \frac{\hbar ^{2}}{2I_{a}};B\equiv \frac{\hbar
^{2}}{2I_{b}};C\equiv \frac{\hbar ^{2}}{2I_{c}}
\end{equation}
and polarizability components parallel, $\alpha _{\parallel }$, and perpendicular, $\alpha _{\perp }$, to the figure axis.

\begin{table}[h]
\centering
\begin{tabular}{c|c|c|c}
\hline\hline
\text{case (i)} \T  & \text{case (ii)} & \text{case (iii)} & \text{case (iv)} \\
\textbf{I}\text{ prolate, } & \textbf{I} \text{ prolate, } &
\textbf{I}\text{
oblate, } & \textbf{I}\text{ oblate, }\\
$\boldsymbol{\alpha }$\ \text{ prolate } \B & $\boldsymbol{\alpha
}$\ \text{ prolate} & $\boldsymbol{\alpha }$\ \text{ oblate} &
$\boldsymbol{\alpha }$\ \text{ oblate}
\\ \hline
$I_{a}\!<\!I_{b}\!=\!I_{c}$ \T & $I_{a}\!<\!I_{b}\!=\!I_{c}$ &
$I_{a}\!=\!I_{b}\!<\!I_{c}$ & $
I_{a}\!=\!I_{b}\!<\!I_{c}$ \\
$A\!>\!B\!=\!C$ & $A\!>\!B\!=\!C$ & $A\!=\!B\!>\!C$ & $A\!=\!B\!>\!C$ \\
$\alpha _{a}\!<\!\alpha _{b}\!=\!\alpha _{c}$ & $\alpha
_{a}\!>\!\alpha _{b}\!=\!\alpha _{c}$ & $\alpha _{a}\!=\!\alpha
_{b}\!>\!\alpha _{c}$ & $\alpha _{a}\!=\!\alpha _{b}\!<\!\alpha
_{c}$ \\
$\alpha _{\parallel }\!<\!\alpha _{\perp }$ \B & $\alpha
_{\parallel }\!>\!\alpha _{\perp }$ & $\alpha _{\parallel
}\!<\!\alpha _{\perp }$ & $\alpha _{\parallel }\!>\!\alpha
_{\perp }$ \\ \hline\hline
\end{tabular}
\caption{\label{tbl:SymComb}Symmetry combinations of the inertia and polarizability tensors for a polarizable symmetric top molecule. See text.}
\end{table}

The rotational Hamiltonian, $H_{r}$, of a symmetric top molecule is given by
\begin{equation}
H_{r}=B\textbf{\textit{J}}^{2}+\rho B J_{z}^{2}  \label{hamrot}
\end{equation}
with $\textbf{\textit{J}}$ the angular momentum operator, $J_{z}$ its projection on the figure axis $z$ ($z\equiv a$ for prolate and $z\equiv c$ for oblate top), and
\begin{equation}
\rho \equiv\left\{\begin{array}{l}
\left(\frac{A}{B}-1\right)>0\textnormal{\hspace{0.2in}for }\mathbf{I}\textnormal{ prolate} \\
\left(\frac{C}{B}-1\right)<0\textnormal{\hspace{0.2in}for }\mathbf{I}\textnormal{ oblate}
\end{array}\right.
\end{equation}

The symmetric-top molecule is subject to a combination of a \emph{static electric field}, $\varepsilon _{S}$, with a \emph{nonresonant laser field}, $ \varepsilon _{L}$. The fields $\varepsilon _{S}$ and $\varepsilon _{L}$ can be tilted with respect to one another by an angle, $\beta $. We limit our consideration to a pulsed plane wave radiation of frequency $\nu$ and time profile $g(t)$ such that
\begin{equation}
\varepsilon _{L}^{2}(t)=\frac{8\pi }{c}Ig(t)\cos ^{2}(2\pi \nu
t)
\end{equation}
where $I$ is the peak laser intensity (in CGS units). We assume the oscillation frequency $ \nu $ to be far removed from any molecular resonance and much higher than either $\tau ^{-1}$ (with $\tau $ the pulse duration) or the rotational periods. The resulting effective Hamiltonian, $H(t)$, is thus averaged over the rapid oscillations. This cancels the interaction between $\mu $ and $ \varepsilon _{L}$ (see also ref. \cite{Dion1999,Henriksen1999}) and reduces the time dependence of $\varepsilon _{L}$ to that of the time profile,

\begin{equation}
\langle \varepsilon _{L}^{2}(t)\rangle =\frac{4\pi }{c}Ig(t)
\end{equation}

Thus the Hamiltonian becomes
\begin{equation}
H(t)=H_{r}+V_{\mu }+V_{\alpha }(t)  \label{ham1}
\end{equation}
where the permanent, $V_{\mu }$, and induced, $V_{\alpha }$, dipole potentials are given by
\begin{eqnarray}
V_{\mu } &=&-B\omega \cos \theta _{S} \\
V_{\alpha }(t) &=&-B\Delta \omega (t)\cos ^{2}\theta
_{L}-B\omega _{\perp }(t)
\end{eqnarray}
with the dimensionless interaction parameters defined as follows
\begin{subequations}
\begin{eqnarray}
\omega &\equiv &\frac{\mu \varepsilon _{S}}{B}   \\
\omega _{||,\bot }(t) &=&\omega _{||,\bot }g(t)   \\
\omega _{||,\bot } &\equiv &\frac{2\pi \alpha _{||,\bot }I}{Bc}   \\
\Delta \omega &\equiv &\omega _{||}-\omega _{\bot }   \\
\Delta \omega (t) &=&\omega _{||}(t)-\omega _{\bot }(t)\equiv
\Delta \omega g(t)
\end{eqnarray}
\end{subequations}

The time-dependent Schr\"{o}dinger equation corresponding to Hamiltonian (\ref{ham1}) can be cast in a dimensionless form
\begin{equation}
i\frac{\hbar }{B}\frac{\partial \psi (t)}{\partial
t}=\frac{H(t)}{B}\psi (t) \label{schr1}
\end{equation}
which clocks the time in units of $\hbar /B$, thus defining a ``short'' and a ``long'' time for any molecule and pulse
duration. In what follows, we limit our consideration to the adiabatic regime, which arises for $\tau \gg \hbar /B$. This is
tantamount to $g(t)\rightarrow 1$, in which case the Hamiltonian (\ref{ham1}) can be written as
\begin{equation}
\frac{H(t)}{B}=\frac{H_{r}}{B}-\omega \cos \theta _{S}-\left(
\Delta \omega \cos ^{2}\theta _{L}+\omega _{\bot }\right)
\label{adham}
\end{equation}
Hence our task is limited to finding the eigenproperties of Hamiltonian (\ref{adham}). For $\Delta\omega=\omega=0$, the
eigenproperties become those of a field-free rotor; the eigenfunctions then coincide with the symmetric-top
wavefunctions, $|JKM\rangle $, and the eigenvalues become $E_{JKM}/B$, with $ K$ and $M$ the projections of the rotational
angular momentum $J$ on the figure and space-fixed axis, respectively. In the high-field limit, $\Delta \omega\rightarrow
\pm \infty $ and/or $\omega \rightarrow \infty $, the range of the polar angle is confined near the quadratic potential minimum, and eq. (\ref{adham}) reduces to that for a two-dimensional angular harmonic oscillator (\emph{harmonic librator}), see Subsection \ref{subsec:Correlation}.

\begin{figure}[htbp]
\centering
\includegraphics[]{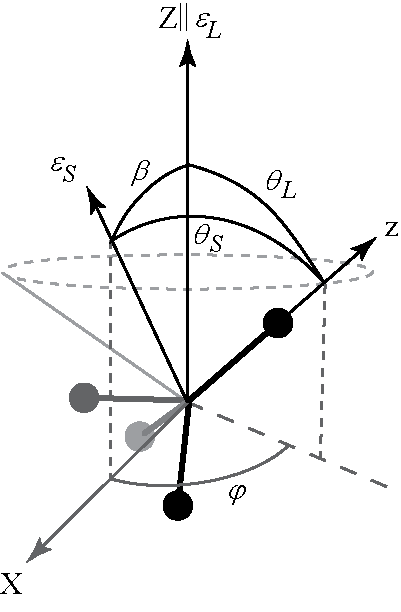}
\caption{
Illustration of the angles used in equation \ref{eqn:AdditionTheorem} for substitution for arbitrary field directions.\label{fig:Geometry}}
\end{figure}

If the tilt angle $\beta $ between the field directions is nonzero, the relation
\begin{equation}  \label{eqn:AdditionTheorem}
\cos \theta _{S}=\cos \beta \cos \theta +\sin \beta \sin \theta
\cos \varphi
\end{equation}
is employed in Hamiltonian (\ref{ham1}), with $\theta
\equiv \theta _{L}$ and $\varphi \equiv \varphi_L$, see Figure \ref{fig:Geometry}.

If the static and radiative fields are \emph{collinear} (i.e., $\beta =0$ or $\pi $), $M$ and $K$ remain good quantum numbers. Note that except when $K=0$, all states are doubly degenerate. While the permanent dipole interaction $ V_{\mu }$ mixes states with $\Delta J=\pm 1$ (which have opposite parities), the induced dipole interaction mixes states with $\Delta J=0\wedge \pm 2$ (which have same parities), but, when $MK\neq 0,$ also states with $\Delta J=\pm 1$ (which have opposite parities). Thus \emph{either} field has the ability to create \emph{oriented states} of mixed parity.

If the static and radiative fields are \emph{not} collinear (i.e., $\beta \neq 0$ or $\pi $), the system no longer possesses cylindrical symmetry. The $\cos \varphi $ operator mixes states which differ by $\Delta M=\pm 1$ and so $ M $ ceases to be a good quantum number.

\begin{figure}[htbp]
\centering
\includegraphics[]{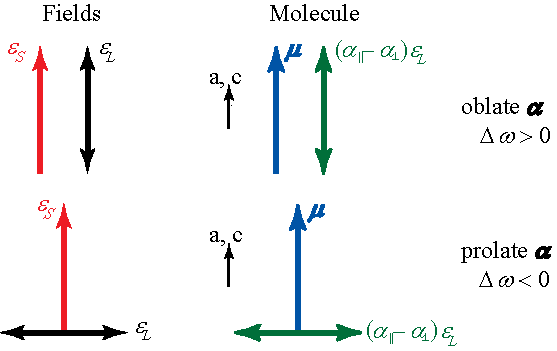}\caption{
(color online) Schematic of the configurations of the fields and dipoles. An electrostatic, $\varepsilon_{S} $, and a linearly
polarized radiative, $\varepsilon _{L}$, field are considered to be either collinear or perpendicular to one another. While the
permanent dipole, $\mu $, of a symmetric top molecule is always along the figure axis ($a$ or $c$ for a prolate or oblate tensor
of inertia), the induced dipole moment, ($\alpha _{||}-\alpha_{\bot })\varepsilon _{L}$, is directed predominantly along the
figure axis for an oblate anisotropy of the polarizability tensor, $\alpha _{||}>\alpha _{\bot }$, and perpendicular to it
for a prolate polarizability, $\alpha _{||}<\alpha _{\bot }$. See Table \ref{tbl:SymComb} and text.\label{fig:FieldsMolecules}}
\end{figure}

A schematic of the field configurations and molecular dipole moments, permanent and induced, is shown is Figure
\ref{fig:FieldsMolecules}.

The elements of the Hamiltonian matrix in the symmetric-top basis set $|JKM\rangle$ possess the following symmetries
\begin{align}
\left\langle J^{\prime }KM^{\prime }\!\left|H\right|\!JKM\right\rangle &=\nonumber\\
(-1)^{M^{\prime }-\!M}& \!\left\langle
J^{\prime }-\!\!K-\!\!M^{\prime }\!\left|H\right|\!J-\!\!K-\!\!M\right\rangle  \label{elem1}
\\
\left\langle J^{\prime }-\!\!KM^{\prime } \!\left| H\right|\!J-\!\!KM\right\rangle &=\nonumber\\ (-1)^{M^{\prime }-M}&\!\left\langle
J^{\prime }K-\!\!M^{\prime }\!\left|H\right|\!JK-\!\!M\right\rangle  \label{elem2}
\end{align}
Since the Hamiltonian has the same diagonal elements for the two symmetry representations belonging to $+K$ and $-K$, it follows that they are connected by a unitary transformation, $U$. On the other hand, complex conjugation, $\mathcal{K}$, of a symmetric top state yields
\begin{equation}
\mathcal{K}\left| JKM\right\rangle =(-1)^{M-K}\left|
J-\!K-\!M\right\rangle
\end{equation}
and so we see from eqs. (\ref{elem1}) and (\ref{elem2}) that the $+K$ and $-K $ representations are related by the combined operation $U\mathcal{K}$, which amounts to time reversal. The two representations have the same eigenenergies and are said to be separable-degenerate \cite{BunkerJensen98}. We note that the $+K$ and $-K$ representations are connected by time-reversal both in the absence and presence of an electric field (see also ref. \cite{Tinkham}); the time reversal of a given matrix element is effected by a multiplication by $(-1)^{M^{\prime}-M}$. The symmetry properties, eqs. (\ref{elem1}) and (\ref{elem2}), are taken advantage of when setting up the Hamiltonian matrix. In what follows, we concentrate on the case of collinear (i.e., $\beta =0$ or $\pi $) and perpendicular fields (i.e., $\beta =\pi /2$).

We label the states in the field by the good quantum number $K$ and the nominal symbols $\tilde{J}$ and $\tilde{M}$ which designate the values of the quantum numbers $J$ and $M$ of the free-rotor state that adiabatically correlates  with the state at $\Delta\omega\neq0$ and/or $\omega>0$, $J\rightarrow \tilde{J}$ and $M\rightarrow \tilde{M}$. We condense our notation by taking $K$ to be nonnegative, but keep in mind that each state with $K\neq 0$ is doubly degenerate, on account of the $+K$ and $-K$ symmetry representations. For collinear fields, $\tilde{M}=M$.

In the \emph{tilted fields}, the adiabatic label of a state depends on the order in which the parameters $\omega,$
$\Delta\omega$ and $\beta $ are turned on, see Subsection \ref{subsec:PerpendicularFields}. As a result, we distinguish among
$\left|\tilde{J},K, \tilde{M};\omega ,\Delta\omega ,\beta \right\rangle$, $\left|\tilde{J},K,\tilde{M} ;\Delta \omega ,\omega ,\beta \right\rangle$,
$\left|\tilde{J},K,\tilde{M};\omega ,\beta,\Delta \omega \right\rangle$, and $\left|\tilde{J},K,\tilde{M};\Delta \omega ,\beta ,\omega \right\rangle
$ states, or $\left|\tilde{J},K,\tilde{M};\{p\}\right\rangle $ for short, with $\{p\}$ any one of the parameter sets $\omega ,\Delta
\omega ,\beta $ or $\Delta \omega ,\omega ,\beta $ or $\omega,\beta ,\Delta \omega $ or $\Delta \omega ,\beta ,\omega $.

The $\left|\tilde{J},K,\tilde{M};\{p\}\right\rangle $ states are recognized as coherent linear superpositions of the field-free symmetric-top states,
\begin{equation}
\left|\tilde{J},K,\tilde{M};\{p\}\right\rangle
=\sum_{J,M}a_{JM}^{\tilde{J}K\tilde{M}}\left| J,K,M\right\rangle
\end{equation}
with $a_{JM}^{\tilde{J}K\tilde{M}}$ the expansion coefficients. For a given state, these depend solely on $\{p\}$,
\begin{equation}
a_{JM}^{\tilde{J}K\tilde{M}}=a_{JM}^{\tilde{J}K\tilde{M}}\left(\{p\}\right)
\end{equation}

The orientation and the alignment are given by the direction-direction (two-vector) correlation \cite{BiedenharnRose} between the dipole moment (permanent or induced) and the field vector ($\varepsilon _{S}$ or $ \varepsilon _{L}$). A direction-direction correlation is characterized by a single angle, here by the polar angle $\theta_{S}$ (for the orientation of the permanent dipole with respect to $\varepsilon _{S}$) or $\theta _{L}$ (for the alignment of the induced dipole with respect to $\varepsilon _{L}$). The distribution in either $\theta _{S}$ or $\theta _{L}$ can be described in terms of a series in Legendre polynomials and
characterized by Legendre moments. The first odd Legendre moment of the distribution in $\theta _{S}$ is related to the
expectation value of $\cos \theta _{S}$, the \emph{orientation cosine},

\begin{eqnarray}
\left\langle \cos \theta _{S}\right\rangle &\equiv&\sum_{J^{\prime
}M^{\prime }}\sum_{JM}a_{J^{\prime }M^{\prime
}}^{\tilde{J}K\tilde{M}*}a_{JM}^{\tilde{J}
K\tilde{M}}(2J+1)^{\frac{1}{2}}(2J^{\prime }+1)^{\frac{1}{2}
} \nonumber\\
&& \times(-1)^{M-K}\!\left(\!
\begin{array}{ccc}
J & 1 & J^{\prime } \\
-K & 0 & K
\end{array}
\right)\nonumber \\
&&\times\left[\! \cos \beta \!\left(
\begin{array}{ccc}
J & 1 & J^{\prime } \\
-M & 0 & M^{\prime }
\end{array}
\right) +\sin \beta \sqrt{\frac{1}{2}}\right. \nonumber\\ &&\left.\times\left( \!\left(
\begin{array}{ccc}
J & 1 & J^{\prime } \\
-M & -1 & M^{\prime }
\end{array}
\right)\! -\!\left(
\begin{array}{ccc}
J & 1 & J^{\prime } \\
-M & 1 & M^{\prime }
\end{array}
\right)\!\right)\! \right]  \nonumber \\
&=&\left\langle\left. \tilde{J},K,\tilde{M};\{p\}\right|\cos \theta
_{S}\left|\tilde{J},K,\tilde{M} ;\{p\}\right.\right\rangle
\end{eqnarray}
and the first even Legendre moment of the distribution in $\theta _{L}$ to the expectation value of $\cos ^{2}\theta_{L}$, the \emph{alignment cosine},

\begin{eqnarray}
\left\langle \cos ^{2}\theta _{L}\right\rangle &\equiv& \sum_{J^{\prime}M} \sum_{J}a_{J^{\prime}M}^{\tilde{J}K\tilde{M}*}a_{JM}^{\tilde{J}K\tilde{M}
}\left[\delta _{JJ^{\prime}}\frac{1}{3}+\frac{2}{3}(2J+1)^{\frac{1}{2}}\right. \nonumber\\
&&\times\!(2J^{\prime }\!+\!1)^{\frac{1}{2}}\!(-\!1)^{M\!-\!K}\left.\!\!\left(
\begin{array}{ccc}
J & 2 & J^{\prime } \\
-M & 0 & M
\end{array}
\right)\!\! \left(
\begin{array}{ccc}
J & 2 & J^{\prime } \\
-K & 0 & K
\end{array}
\right)\! \right]  \nonumber \\
&=&\left\langle\left. \tilde{J},K,\tilde{M};\{p\}\right|\cos ^{2}\theta _{L}\left|\tilde{J},K,\tilde{M};\{p\}\right.\right\rangle
\end{eqnarray}
The states with same $|K|$ and same $MK$ have both the same
energy,
\begin{equation}
E\left(\tilde{J},+\!K,\tilde{M};\{p\}\right)=E\left(\tilde{J},-\!K,-\!\tilde{M};\{p\}\right)
\end{equation}
and the same directional properties, as follows from the Hellmann-Feynman theorem,
\begin{equation}
\langle \cos \theta _{S}\rangle =-\frac{\partial \left(
\frac{E}{B}\right) }{\partial \omega }
\end{equation}
and
\begin{equation}
\left\langle \cos ^{2}\theta _{L}\right\rangle =-\frac{\partial \left(
\frac{E}{B} +\omega _{\bot }\right) }{\partial \Delta \omega }
\label{theta2}
\end{equation}
The angular amplitudes of the permanent and induced dipoles are given, respectively, by

\begin{equation}
\theta _{S,0}=\arccos \langle \cos \theta _{S}\rangle
\label{eqn:thetas0}
\end{equation}
and

\begin{equation}
\theta _{L,0}=\arccos \left[\langle \cos ^{2}\theta _{S}\rangle\right
]^{\frac{1}{2}} \label{eqn:thetal0}
\end{equation}

The elements of the Hamiltonian matrix, evaluated in the symmetric-top basis set, are listed in Appendix \ref{MatrixElements}. The dimension of the Hamiltonian matrix determines the accuracy of the eigenproperties computed by diagonalization. For the collinear case, the dimension of the matrix is given as $J_{\max }+1-\max (|K|,|M|)$, with $J$ ranging between $\max (|K|,|M|)$ and $J_{\max}$. For the states and field strengths considered here, a $12\times 12$ matrix yields an improvement of less than 0.1\% of $E/B$ over a $11\times 11$ matrix. But the convergence depends strongly on the state considered. For the perpendicular case, the dimension of the matrix is $(J_{max}+1)^{2}-K^{2}$ with $J$ ranging between $|K|$ and $J_{max}$ and $M$ between $-J$ and $J$. Convergence within 0.1\% can be achieved for the states considered with $J_{max}=10$.

The $\left|\tilde{J},K,\tilde{M};\{p\}\right\rangle$ states are not only labeled but also identified in our computations by way of their
adiabatic correlation with the field-free states. The states obtained by the diagonalization procedure cannot be identified
by sorting of the eigenvalues, especially for perpendicular fields when the states undergo numerous crossings, genuine as
well as avoided. Therefore, we developed an identification algorithm based on a gradual perturbation of the field-free
symmetric-top states. Instead of relying on the eigenvalues, we compare the wavefunctions, which are deemed to belong to the
same state when their coefficients evolve continuously through a crossing as a function of the parameters $\{p\}$. If $\left| \Psi _{0}\right\rangle $ is the state to be tracked, one has to calculate its overlap $\left\langle \Psi _{k}|\Psi
_{0}\right\rangle $ with all the eigenvectors $\left| \Psi _{k}\right\rangle $ that pertain to the Hamiltonian matrix with
the incremented parameters. The state with the largest overlap is then taken as the continuation of $\Psi _{0}$. This method
has been used for the general problem of tilted fields.

\section{Effective potential\label{sec:EffPot}}

In order to obtain the most probable spatial distribution (geometry), wavefunctions, such as

\begin{equation}
\left|\tilde{J},K,\tilde{M};\{p\}\right\rangle =\Xi _{\tilde{J},K,\tilde{M}
;\{p\}}(\theta ,\varphi )\equiv \Xi (\theta ,\varphi )
\end{equation}
obtained by solving the Schr\"{o}dinger equation in curvilinear coordinates, need to be properly spatially weighted by a non-unit Jacobian factor, here $ \sin \theta $. Alternatively, a wavefunction,
\begin{equation}
|\Phi |^{2}=|\Xi |^{2}\sin \theta
\end{equation}
with a \emph{unit Jacobian} can be constructed which gives the most probable geometry directly; such a wavefunction is an
eigenfunction of a Hamiltonian
\begin{equation}
\frac{H^{\prime }}{B}=-\frac{d^{2}}{d\theta ^{2}}+U
\label{Hprime}
\end{equation}
where $U$ is an effective potential \cite{FriedHerschb99-1,FriedHerschb99-2}. Eq. (\ref{Hprime}) shows that $\Phi $ corresponds to the solution of a 1-D
Schr\"{o}dinger equation for the curvilinear coordinate $\theta $ and for the effective potential $U$. Since $ \Phi $ can only
take significant values within the range demarcated by $U$, one can glean the geometry from the effective potential and the
eigenenergy. In this way, one can gain insight into the qualitative features of the eigenproblem without the need to
find the eigenfunctions explicitly. Conversely, one can use the concept of the effective potential to organize the solutions and
to interpret them in geometrical terms.

For collinear fields, the effective potential takes the form
\begin{eqnarray}
U &=&\left[ \frac{M^{2}-\frac{1}{4}}{\sin ^{2}\theta
}-\frac{1}{4}\right]
-\rho K^{2}+\frac{K^{2}-2MK\cos \theta }{\sin ^{2}\theta }  \nonumber \\
&&-\omega \cos \theta -\left( \Delta \omega \cos ^{2}\theta
+\omega _{\bot }\right)  \label{eqn:effpot}
\end{eqnarray}
Its first term, symmetric about $\theta =\pi /2$, arises due to the centrifugal effects and, for $|M|>0$, provides a repulsive
contribution competing with the permanent (fourth term) and induced (fifth term) interactions. For $|K|>0$, the second term
just uniformly shifts the potential, either down when $\rho >0$ (prolate top), or up when $\rho <0$ (oblate top). The third term
provides a contribution which is \emph{asymmetric} with respect to $\theta =\pi /2$ for \emph{precessing states}, i.e., states with $MK\neq 0$. It is this term which is responsible for the first-order Stark effect in symmetric tops
and for the inherent orientation their precessing states possess. The fourth term, due to the permanent dipole
interaction, is asymmetric with respect to $\theta =\pi /2$ for any state, and accounts for all higher-order Stark effects. The
fifth, induced-dipole term, is symmetric about $\theta =\pi /2$. However, it gives rise to a single well for $\boldsymbol{\alpha}
$ prolate ($\alpha_{\parallel }<\alpha_{\perp }$) and a double-well for $\boldsymbol{\alpha} $ oblate ($\alpha
_{\parallel }>\alpha _{\perp }$). This is of key importance in determining the energy level structure and the directionality of
the states bound by the wells.

\section{Behavior of the eigenstates\label{sec:Disc}}

\subsection{Correlation diagrams\label{subsec:Correlation}}

\begin{figure*}[htbp]
\centering
\includegraphics[width=0.85\textwidth]{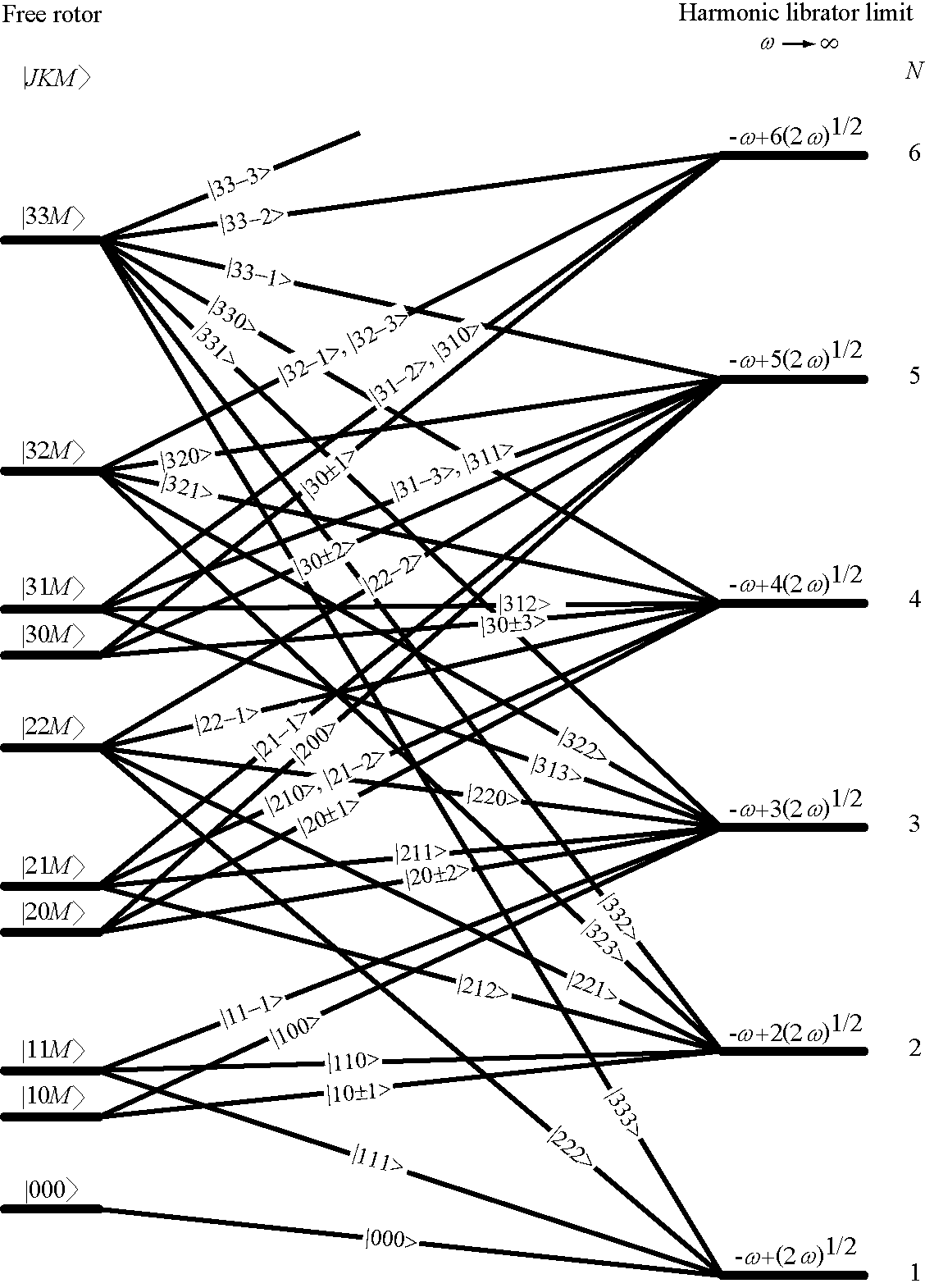}
\caption{Correlation diagram, for the permanent dipole interaction, between the field-free ($\omega \rightarrow 0$) symmetric top states $\left|JKM\right\rangle$ and the harmonic librator states $\left|N\right\rangle$ obtained in the high-field limit ($\omega\rightarrow \infty $), see also \cite{frislenher}. At intermediate fields, the states are labeled by $\left|\tilde{J}KM\right\rangle$.\label{fig:CorrelationElectric}}
\end{figure*}

\begin{figure*}[htbp]
\centering
\includegraphics[width=0.85\textwidth]{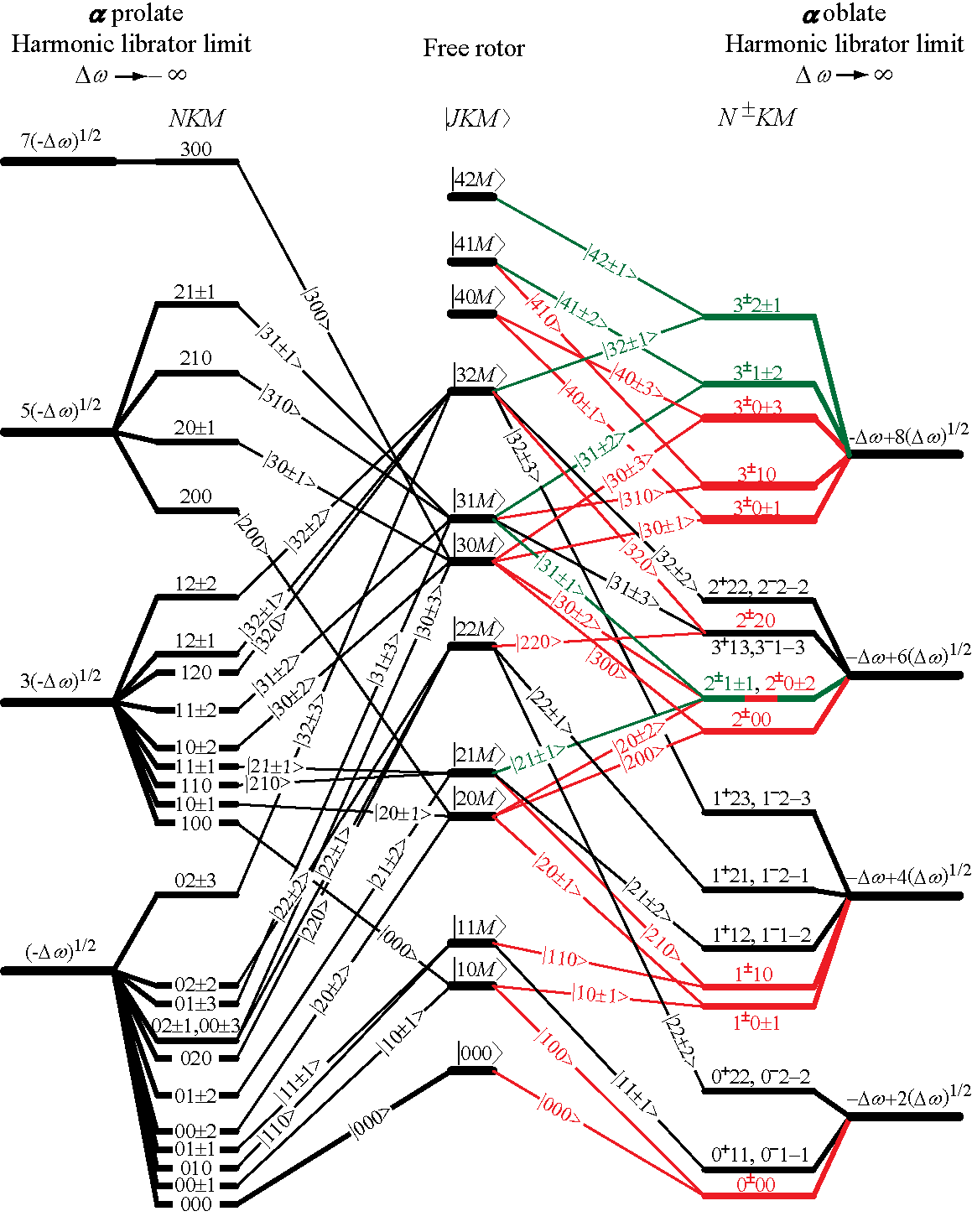}
\caption{(color online) Correlation diagram, for the induced-dipole interaction, between the field-free ($\Delta \omega\rightarrow0$) symmetric top states $\left|JKM\right\rangle$ (center) and the harmonic librator states $\left|NKM\right\rangle$ obtained in the high-field limit for the prolate, $\Delta \omega \rightarrow -\infty $ (on the left), and oblate, $\Delta \omega \rightarrow \infty $ (on the right) case. The harmonic librator states are labeled by the librator quantum number $N$ and the projection quantum numbers $K$ and $M$. At intermediate fields, the states are labeled by $\left| \tilde{J}KM \right\rangle $. States that form tunneling doublets (only for $\Delta\omega\rightarrow\infty$) have $\tilde{J}\geq |K|+|M|$ and are shown in color: red for doublets with $KM=0$, green for doublets with $K,M\neq 0.$ Members of the degenerate doublets have $\tilde{J}<|K|+|M|$ and are shown in black. See text.\label{fig:CorrelationLaser}}
\end{figure*}

In the strong-field limit, a symmetric top molecule becomes a harmonic librator whose eigenproperties can be obtained in
closed form. The eigenenergies in the harmonic-librator limit are listed in Tables \ref{tbl:HarmLibrElectric} and
\ref{tbl:HarmLibrLaser} and used in constructing the correlation diagrams between the field-free and the harmonic librator
limits, shown in Figures \ref{fig:CorrelationElectric} (for the permanent dipole interaction, $\omega \rightarrow \infty $) and
\ref {fig:CorrelationLaser} (for the prolate, $\Delta \omega \rightarrow -\infty $, and oblate, $\Delta \omega \rightarrow
\infty $, induced-dipole interaction).

\begin{table}[hbtp]
\centering
\begin{tabular}{rl}
\hline\hline $\omega \T \B \rightarrow \infty $ \\
\hline 
$\frac{E_{N,K,M}}{B}=$\T &$\rho
K^{2}+N(2\omega )^{\frac{1}{
2}}+KM $\\ & $+\frac{1}{8}\left[ 3(K-M)^{2}-3-N^{2}\right]-\omega$ \\
 $N=$\B & $2\tilde{J}-|K+M|+1$ \\ \hline\hline
\end{tabular} 
\caption{\label{tbl:HarmLibrElectric}Eigenenergies for the permanent dipole interaction in the harmonic librator limit. See \cite{ROST1992,MAERGOIZ1993}.}
\end{table}

\begin{table*}[htbp]
\centering
\begin{tabular}{c|c}
\hline\hline $\boldsymbol{\alpha }$ prolate: $\Delta \omega
\rightarrow -\infty $ \T \B & $\boldsymbol{
\alpha }$ oblate: $\Delta \omega \rightarrow \infty $ \\
\hline $\frac{E_{N,K,M}}{B}=\rho K^{2}+$ \T &
$\frac{E_{N^{\pm
},K,M}}{B}=\rho K^{2}+$ \\
$+(2N+1)(-\Delta \omega )^{\frac{1}{2}}+M^{2}+K^{2}-$ &
$+2(N^{\pm }+1)(\Delta \omega
)^{\frac{1}{2}}+\frac{M^{2}}{2}+\frac{K^{2}}{2}-\frac{
(N^{\pm }+1)^{2}}{2}-$ \\
$-\frac{1}{4}(2N^{2}+2N+3)-\omega _{\bot }$ & $-\Delta \omega
-\omega _{\bot
}-\frac{1}{2}$ \\
&  \\
$N=\tilde{J}-|M|$ for $|M|\geq |K|$ & for $\tilde{J}<|M|+|K|$ \\
$N=\tilde{J}-|K|$ for $|M|<|K|$ & $N^{\pm }=2\tilde{J}-|K|-|M|$
for $
KM \gtrless 0$ \\
&  \\
& for $\tilde{J}\geq |M|+|K|$ \\
& $N^{\pm }=\tilde{J}-1$ for $KM\gtrless 0$ when $\tilde{J}-|K+M|$ is odd \\
& $N^{\pm }=\tilde{J}$ for $KM\gtrless 0$ when $\tilde{J}-|K+M|$ is even \\
&  \\
& for $K$ or $M=0$ \\
& $N^{-}=\tilde{J}-1$ when $\tilde{J}-|M|$ or $\tilde{J}-|K|$ is odd \\
\B & $N^{+}=\tilde{J}$ when $\tilde{J}-|M|$ or $\tilde{J}-|K|$ is even \\
\hline\hline
\end{tabular}
\caption{\label{tbl:HarmLibrLaser}Eigenenergies for the induced dipole interaction in the harmonic librator limit. See also
\cite{KimFelker98-1}.}
\end{table*}

The correlation diagram for the permanent dipole interaction, Fig. \ref{fig:CorrelationElectric}, reveals that states with
$K=0$ split into $ \tilde{J}+1$ doublets, each with the same value of $|M|$. The other states, on the other hand, split into
$\tilde{J}+|K|$ at least doubly degenerate states, each of which is characterized by a value of $|K+M|$ for a given $\tilde{J}$. For $|K|<\tilde{J}$ some states are more than doubly degenerate. In the harmonic librator limit, the levels are \emph{infinitely
degenerate} and are separated by an energy difference of $(2\omega )^{1/2}$.

The correlation diagrams for the induced dipole interaction reveal that states with $\tilde{J}<|M|+|K|$ (shown by black
lines in Fig. \ref{fig:CorrelationLaser}) for $\boldsymbol{\alpha }$ oblate and \emph{all} states for
$\boldsymbol{\alpha }$ prolate which have same $|MK|$ form \emph{degenerate doublets}. In contradistinction, states with
$\tilde{J}\geq |M|+|K|$ (shown by red and green lines in Fig. \ref{fig:CorrelationLaser} for $K=0$ and $ K\neq 0$, respectively) bound by $V_{\alpha }$ oblate occur as \emph{tunneling doublets. }This behavior reflects a crucial
difference between the $\boldsymbol{\alpha }$ prolate and $\boldsymbol{\alpha }$ oblate case, namely that the
induced-dipole potential, $V_{\alpha }$, is a double well potential for $ \boldsymbol{\alpha }$ oblate and a single-well
potential for $\boldsymbol{\alpha }$ prolate.

The members of a given tunneling doublet have same values of $KM$ and $|K|$, but $\tilde{J}$'s that differ by $\pm 1$. The
tunneling splitting between the members of a given tunneling doublet decreases with increasing $\Delta \omega $ as
$\exp(a-b\Delta \omega ^{1/2})$, with $a,b\geq 0$, rendering a tunneling doublet quasi-degenerate at a sufficiently large field
strength. In the harmonic librator limit, the quasi-degenerate members of a given tunneling doublet coincide with the $N^{+}$
and $N^{-}$ states, see Table \ref{tbl:HarmLibrLaser}. The $N^{+}$ and $N^{-}$ states with $N^{+}=N^{-}$ for $\tilde{J}
<|M|+|K| $ always pertain to the same $\tilde{J}$ but different $KM$ and so are precluded from forming tunneling doublets as
they \emph{do not} interact. In the $\boldsymbol{\alpha }$ prolate case, the formation of any tunneling doublets is barred
by the absence of a double well. Note that in both the prolate and oblate case, the harmonic librator levels are\emph{\
infinitely degenerate} and their spacing is equal to $2|\Delta \omega |^{1/2}$.

The correlation diagrams of Figs. \ref{fig:CorrelationElectric} and \ref{fig:CorrelationLaser} reveal another key difference
between the permanent and induced dipole interactions, namely the ordering of levels pertaining to the same $\tilde{J}$. The
energies of the levels due to $V_{\mu }$ increase with increasing $|K+M|$. For $V_{\alpha}$ prolate, they decrease with
increasing $|M|$ for levels with $|M|\geq|K|$ while states with $|M|\leq|K|$ have the same asymptote. The energy level pattern
becomes even more complex for $V_{\alpha}$ oblate, which gives rise to $|K|+1$ asymptotes. If $|K|=\tilde{J}$, the energy decreases with
increasing $|M|$, while for $|K|<\tilde{J}$ it only decreases for levels with $|M|+|K|\geq\tilde{J}$. All other levels connect
alternately to the asymptotes with $N^{(\pm)}=\tilde{J}$ or $\tilde{J}-1$. This leads to a tangle of crossings, avoided or
not, once the two interactions are combined, as will be exemplified below.

\subsection{Collinear fields}

\begin{figure*}[htbp]
\centering
\includegraphics[width=1.\textwidth]{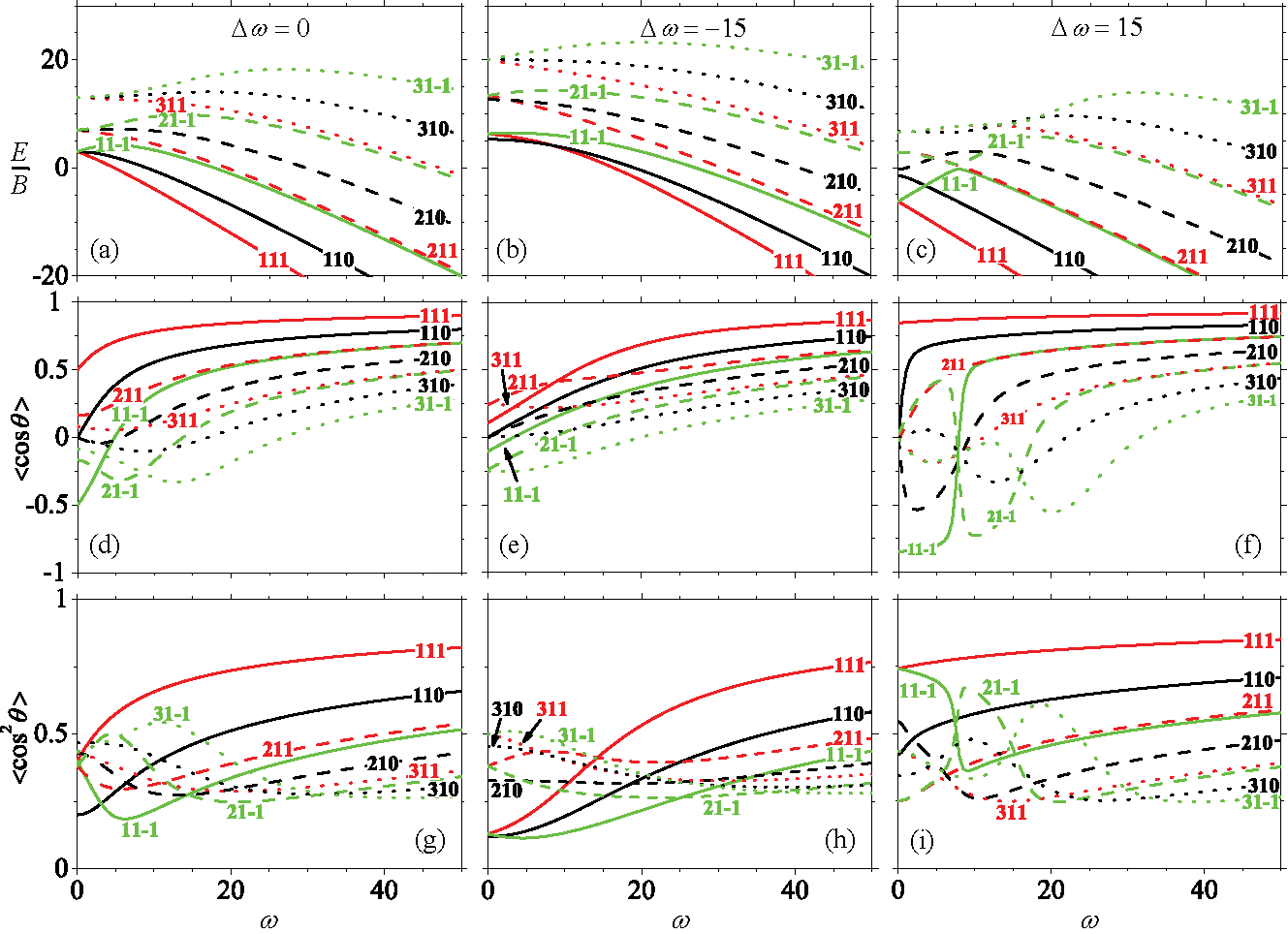}
\caption{(color online) Dependence, in collinear fields,of the eigenenergies, panels (a)-(c), orientation cosines, panels (d)-(f), and alignment cosines, panels (g)-(i), of the states with $0<\tilde{J}\leq 3$, $-1\leq MK\leq 1$, and $K=1$ on the dimensionless parameter $\omega$ (which characterizes the permanent dipole interaction with the electrostatic field) for fixed values of the parameter $\Delta\omega $ (which characterizes the induced-dipole interaction with the radiative field; $\Delta \omega <0$ for prolate polarizability anisotropy, $\Delta \omega >0$ for oblate polarizability anisotropy). The states are labeled by $|\tilde{J}KM\rangle$. Note that panels (a), (d), and (g) pertain to the permanent dipole interaction alone. See text.\label{fig:Electric} }
\end{figure*}

\begin{figure*}[htbp]
\centering
\includegraphics[width=1.\textwidth]{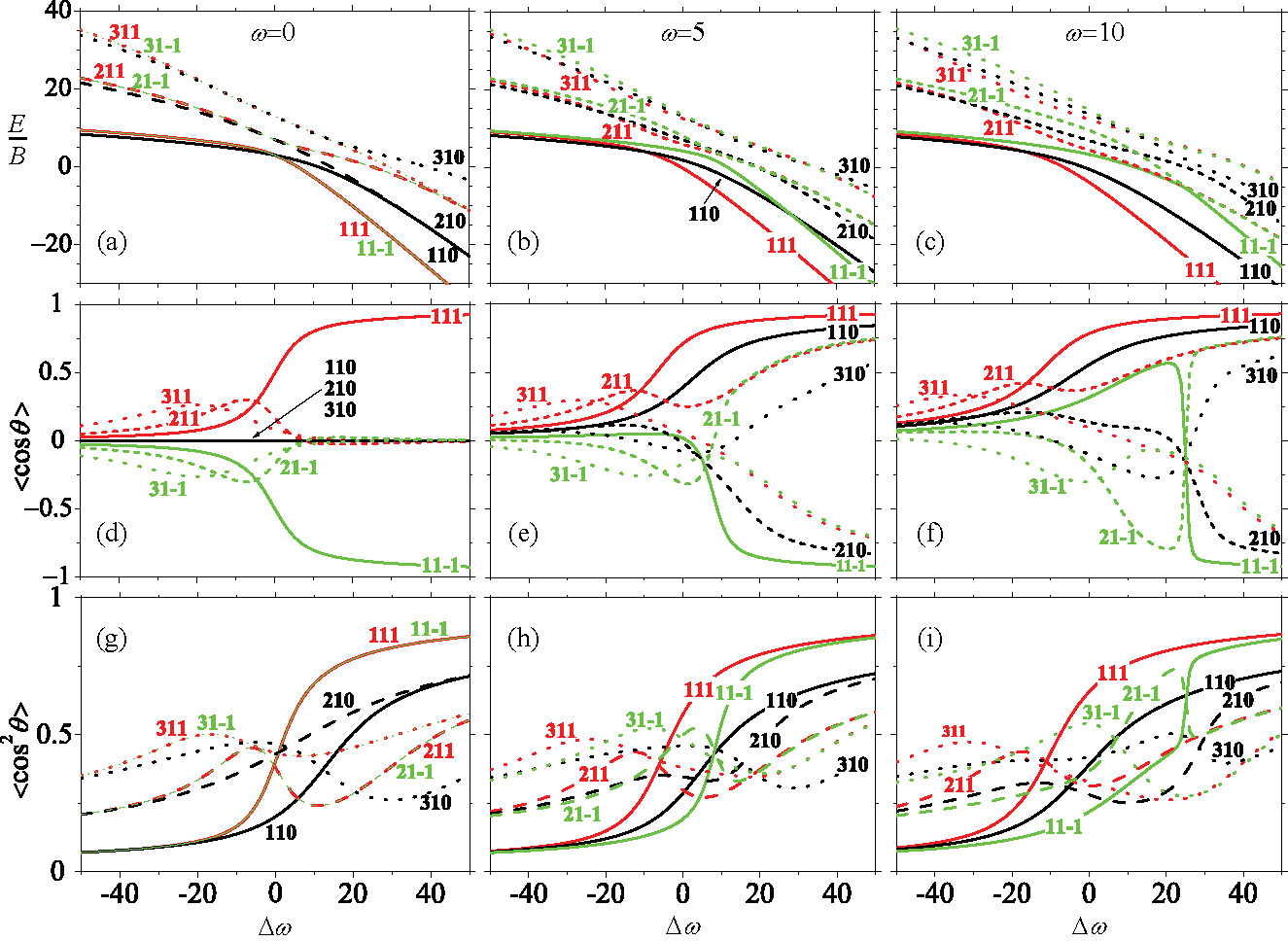}
\caption{(color online) Dependence, in collinear fields, of the eigenenergies, panels (a)-(c), orientation cosines, panels (d)-(f), and alignment cosines, panels (g)-(i), of the states with $0<\tilde{J}\leq 3$, $-1\leq MK\leq 1$, and $K=1$ on the dimensionless parameter $\Delta\omega $ (which characterizes the induced-dipole interaction with the radiative field; $\Delta \omega <0$ for prolate polarizability anisotropy, $\Delta \omega >0$ for oblate polarizability anisotropy) for fixed values of the parameter $\omega $ (which characterizes the permanent dipole interaction with the radiative field). The states are labeled by $|\tilde{J}KM\rangle$. Note that panels (a), (d), and (g) pertain to the induced-dipole interaction alone. See text.\label{fig:Laser} }
\end{figure*}

Figures \ref{fig:Electric} and \ref{fig:Laser} display the dependence of the eigenenergies, panels (a)-(c), orientation cosines, panels (d)-(f), and alignment cosines, panels (g)-(i), of the states with $0<\tilde{J}\leq 3$, $ -1\leq MK\leq 1$, and $K=1$ on the dimensionless parameters $\omega $ and $ \Delta\omega $ that characterize the permanent and induced dipole interactions. These states were chosen as examples since they well represent the behavior of a symmetric top in the combined fields. The two figures show the dependence on $\Delta \omega $ for fixed values of $\omega $ and vice versa. Note that negative values of $\Delta \omega $ correspond to $\boldsymbol{ \alpha }$ prolate and positive values to $\boldsymbol{\alpha }$ oblate. The plots were constructed for $\mathbf{I}$ prolate with $A/B=2$ but, apart from a constant shift, the curves shown are identical with those for $\mathbf{I}$ oblate. Thus Figs. \ref{fig:Electric} and \ref{fig:Laser} represent the entire spectrum of possibilities as classified in Table \ref{tbl:SymComb}.

The left panels of Figs. \ref{fig:Electric} and \ref{fig:Laser} show the eigenenergies and the orientation and alignment cosines
for the cases of pure permanent and pure induced dipole interactions, respectively.

For an angle
\begin{equation}
\gamma \equiv \arccos \frac{KM}{J(J+1)}
\end{equation}
such that $0<\gamma <\pi /2$, the \emph{pure permanent dipole interaction}, $\omega >0$ and $\Delta \omega =0$, panels (a), (d), (g) of Fig. \ref {fig:Electric}, produces states whose eigenenergies decrease with increasing field strength (i.e., the states are\emph{\ high-field seeking}) at all values of $\omega$ and their orientation cosines are positive, which signifies that the body-fixed dipole moment is oriented along $\varepsilon_{S}$ (\emph{right-way orientation}). For states with $\pi/2<\gamma <\pi$, the eigenenergies first increase with $\omega $ (i.e., the states are \emph{low-field seeking}) and the body-fixed dipole is oriented oppositely with respect to the direction of the static field $\varepsilon _{S}$ (the so called \emph{wrong-way orientation}). For states with $K=0$, the angle $\gamma $ becomes the tilt angle of the angular momentum vector with respect to the field direction, which, for $J>0$, is given by
\begin{equation}
\gamma _{0}\equiv \arccos \frac{|M|}{\left[ J(J+1)\right]
^{\frac{1}{2}}}
\end{equation}
At small $\omega $, states with $\gamma_{0}>3^{-1/2}$ are low-field seeking and exhibit the wrong-way orientation, while states with $\gamma_{0}<3^{-1/2}$ are high-field seeking and right-way oriented. 

At large-enough values of $\omega $, all states, including those with $K=0$, become high-field seeking and exhibit right-way orientation. In any case, the dipole has the lowest energy when oriented along the field. Since the asymmetric effective potential (\ref{eqn:effpot}) disfavors angles near $ 180^{\circ}$, and increasingly so with increasing $KM$, the eigenenergies for a given $\tilde{J}$ decrease with increasing $KM$. The ordering of the levels pertaining to the same $\tilde{J}$ is then such that states with higher $KM$ are always lower in energy.

The eigenenergies and orientation and alignment cosines for a \emph{pure induced-dipole interaction}, $\omega =0$ and $|\Delta\omega |>0$, are shown in panels (a), (d), and (g) of Fig. \ref{fig:Laser}. The eigenenergies are given by
\begin{equation}
\frac{E}{B}=\frac{\langle H\rangle }{B}=J(J+1)+\rho K^{2}-\Delta
\omega \langle \cos ^{2}\theta \rangle -\omega _{\bot }
\label{efull}
\end{equation}
cf. eqs. (\ref{hamrot}) and (\ref{adham}). However, Figs. \ref{fig:Laser}a-c and \ref{fig:PF-Laser} only shows $E/B+\omega_{\bot }\equiv \lambda $, which increase with increasing laser intensity for $\boldsymbol{\alpha }$ prolate and decrease for $\boldsymbol{\alpha }$ oblate. Note that both prolate and oblate eigenenergies, $ E/B=\lambda -\omega _{\bot }$, decrease with increasing laser intensity, and so all states created by a pure induced-dipole interaction are high-field seeking as a result. Note that
\begin{equation}
\langle \cos ^{2}\theta \rangle =-\frac{\partial \lambda
}{\partial \Delta \omega }
\end{equation}
and thus the alignment cosines are given by the negative slopes of the curves shown in Fig. \ref{fig:Laser}.

\begin{figure}[htbp]
\centering
\includegraphics[]{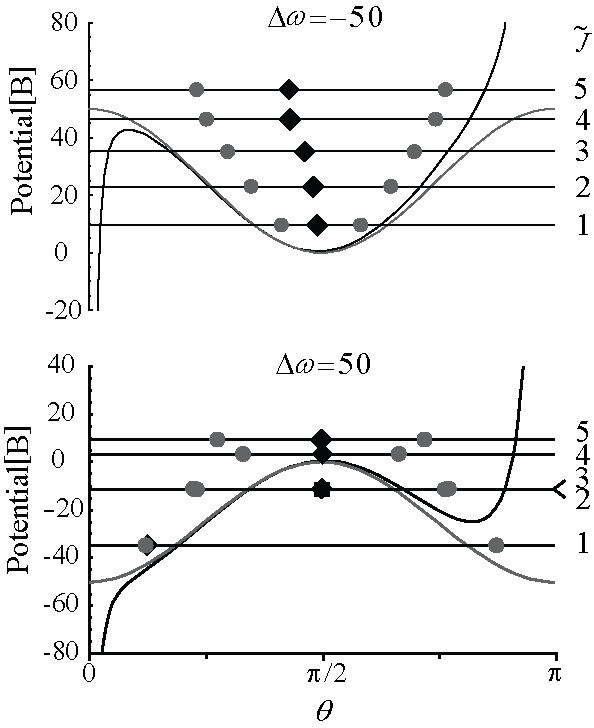}
\caption{Effective potential, $U$, for $KM=1$ and $K=1$ along with the eigenenergies of states with
$\tilde{J}=1,2,...,5$ (horizontal lines) and their alignment (diamonds) and orientation (circles) amplitudes $\theta _{L,0}$ and $\theta _{S,0}$. The grey line shows the induced-dipole potential, $V_{\alpha }$. See eqs. (\ref{eqn:effpot}), (\ref{eqn:thetas0}), (\ref{eqn:thetal0}) and text.\label{fig:EffPotOmegaZero}}
\end{figure}

In panels (d) and (g) one can see that only the precessing states are oriented, and that their orientation shows a dependence on the $\Delta \omega $ parameter which qualitatively differs for $\boldsymbol{\alpha }$ oblate and $\boldsymbol{\alpha }$ prolate: for $\Delta \omega >0$, the orientation is enhanced for states with $\tilde{J}<|M|+|K|$ and suppressed for states with $\tilde{J}\geq |M|+|K|$, while for $\Delta \omega <0$ it tends to be suppressed by the radiative field for all states. This behavior follows readily from the form of the effective potential, eq. (\ref{eqn:effpot}), as shown in Figure \ref{fig:EffPotOmegaZero}. For the prolate case ($ \Delta \omega <0$), the potential becomes increasingly centered at $\theta \rightarrow \pi /2$ with increasing field strength, and therefore tends to force the body-fixed electric dipole (and thus the figure axis) into a direction perpendicular to the field. On the other hand, for the oblate case ($\Delta \omega >0$), the effective potential provides, respectively, a forward ($ \theta \rightarrow 0$) and a backward ($\theta \rightarrow\pi $) well for the $N^{+}$ and $N^{-}$ states of a tunneling doublet (for $\tilde{J}\geq |M|+|K|$) or of a degenerate doublet (for $\tilde{J}<|M|+|K|$). However, only for the degenerate-doublet states does the increasing field strength result in an enhanced orientation at $\varepsilon _{S}=0$. This distinction is captured by the effective potential, Fig. \ref{fig:EffPotOmegaZero}, whose asymmetric forward (for $MK>0$) or backward ($MK<0$) well lures in the $\tilde{J}<|M|+|K|$ states. The $\tilde{J}\geq |M|+|K|$ states become significantly bound by the $V_{\alpha }$ oblate potential at $\Delta \omega$ values large enough to make them feel the double well, which makes the two opposite orientations nearly equiprobable.

\begin{figure}[htbp]
\centering
\includegraphics{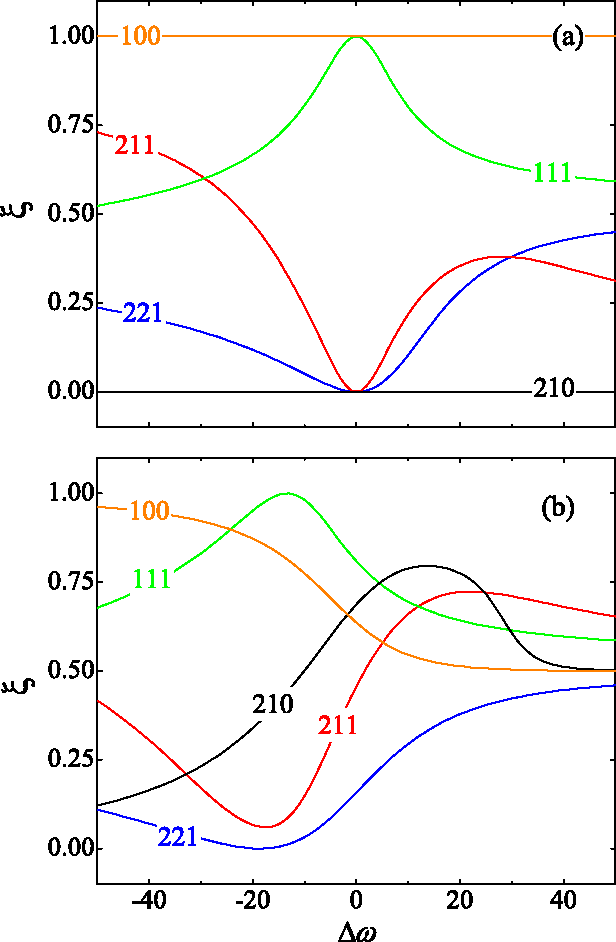}
\caption{(color online) Dependence of the $J$-parity mixing parameter $\xi $ on the $\Delta \omega$ parameter at $\omega =0$ (panel a) and $\omega =10$ (panel b). Note that the better the $J$-parity mixing the closer is the $\xi $ parameter to $1/2.$ See text.\label{fig:Mixing} }
\end{figure}

Figure \ref{fig:Mixing} shows the dependence of the $J$-state parity mixing on the $\Delta \omega $ parameter at $\omega =0$
(panel a) and $\omega =10$ (panel b). The mixing is captured by a parameter
\begin{equation}
\xi \equiv \sum_{J=2n+1}\left(
a_{JM}^{\tilde{J}K\tilde{M}}\right) ^{2}
\end{equation}
with $n$ an integer. In the absence of even-odd $J$ mixing, $\xi=0$ for $ \tilde{J}$ even and $\xi =1$ for $\tilde{J}$ odd; for a ``perfect'' $J$-parity mixing, $\xi =\frac{1}{2}$ for either $\tilde{J}$ even or odd. We see that for the sampling of states shown, the non-precessing states become parity mixed only when $\omega >0$. However, all precessing states are $J$-parity mixed as long as $\Delta \omega \neq 0$.

The eigenenergies as well as eigenfunctions in the harmonic librator limit for both the prolate and oblate polarizability interactions have been found previously by Kim and Felker \cite{KimFelker98-1}, and we made use of the former in constructing the correlation diagram in Fig. \ref{fig:CorrelationLaser}. We note that in the prolate case, the eigenenergies and alignment cosines, as calculated from Kim and Felker's eigenfunctions, agree with our numerical calculations for the states considered within 4\% already at $ \Delta \omega =-50$. The prolate harmonic librator eigenfunctions render, however, the orientation cosines as equal to zero, which they are generally not at any finite value of $\Delta \omega $.

For the $\boldsymbol{\alpha}$ oblate case, the eigenenergies in the harmonic librator limit agree with those obtained numerically for the states considered within 5\% at $\Delta\omega=50$. For $\Delta\omega\approx350$, the difference between the alignment cosines obtained numerically and from  the analytic eigenfunctions is less then 3\%. We note that for $\tilde{J}<|M|+|K|$, only one eigenfunction exists, pertaining either to $N^{+}$ for $KM>0$ or to $N^{-}$ for $KM<0$, cf. Table \ref{tbl:HarmLibrLaser}. This eigenfunction is strongly directional, lending a nearly perfect right-way orientation to an $N^{+}$ state and a nearly perfect wrong-way orientation to an $N^{-}$ state. These eigenfunctions pertain to the degenerate doublets. The orientation cosines can be obtained from the analytic wavefunctions within $0.01$\% for $\Delta\omega=100$. For $\tilde{J}\geq |M|+|K|$ (including the cases when either $K=0$ \textnormal{or} $M=0$) both the $N^{+}$ and $N^{-}$ analytic solutions exist, cf. Table \ref{tbl:HarmLibrLaser}, and pertain to the tunneling doublets. The degeneracy of the doublets in the $\Delta\omega\rightarrow +\infty$ limit precludes using the analytic eigenfunctions in calculating the orientation cosines. In contrast to the numerical results, the analytic solution predicts a strong orientation, which in fact is not present, as shown in Fig. \ref{fig:EffPotOmegaZero}. The linear combination of the analytic eigenfunctions $f^+$ and $f^-$, which is also a solution, given as $f_{1,2}=1/\sqrt{2}(f^+\pm f^-)$, does not exhibit any orientation, just alignment.

The above behavior of symmetric-top molecules as a function of $\Delta \omega $ at $\omega =0$ sets the stage for what happens once the static field is turned on and so $\omega >0$. For $\Delta \omega >0$, the combined electrostatic and radiative fields act \emph{synergistically}, making \emph{all states sharply oriented}. For $\Delta \omega <0$, the orientation either remains nearly zero (for states with $KM=0$) or tends to vanish (for states with $KM\neq 0$) with increasing $|\Delta\omega |$.

The synergistic action of the combined fields arises in two different ways, depending on whether $\tilde{J}<|M|+|K|$ or $\tilde{J}\geq |M|+|K|$.

For $\tilde{J}\geq |M|+|K|$, the orientation is due to a coupling of the members of the tunneling doublets (e.g., the $|2,1,1\rangle $ and $ |3,1,1\rangle $ states) by the permanent dipole interaction. The tunneling doublets occur, and hence this mechanism is in place, for $\Delta \omega >0$. The coupling of the tunneling-doublet members by the permanent dipole interaction is the more effective the smaller is the level splitting between the doublet members (which correlate with the $N^{+}$ and $N^{-}$ in the harmonic librator limit). Since the levels of a tunneling doublet can be drawn arbitrarily close to one another by the induced dipole interaction, the coupling of its members by even a weak electrostatic field can be quite effective, resulting in a strong orientation of both states. The wrong-way orientation of the upper members of the tunneling doublets can be converted into a right-way orientation. Such a conversion takes place at sufficiently large $\omega$ where the permanent dipole interaction prevails over the induced dipole interaction. As noted in previous work \cite{FriedHerschb99-2}, the coupling of the tunneling doublets by $V_{\mu }$ also arises for the non-precessing states, in which case one can speak of a \emph{pseudo-first-order Stark effect} in the combined fields. The precessing states show in addition the well known first-order Stark effect in the electrostatic field alone, which relies on the coupling of states with the same $|K|$ and does not involve any hybridization of $J$.

As noted above, the $\tilde{J}<|M|+|K|$ states are strongly oriented by the induced-dipole interaction alone. Since the members of a degenerate doublet that correlate with the $N^{+}$ and $N^{-}$ states (e.g., the $|1,1,1\rangle $ and $|1,1,-1\rangle $ states) have \emph{different} values of $KM$, adding an electrostatic field does not lead to their coupling, as collinear $V_{\mu }$ and $V_{\alpha }$ can only mix states with same $KM$. However, the static field skews the effective potential $U$, eq. (\ref{eqn:effpot}), that enhances the orientation of the right-way oriented states ($N^{+}$) and, at sufficiently high $\omega $, reverses the orientation of the initially wrong-way oriented states ($N^{-}$), see below.

\begin{figure*}[htbp]
\centering
\includegraphics[]{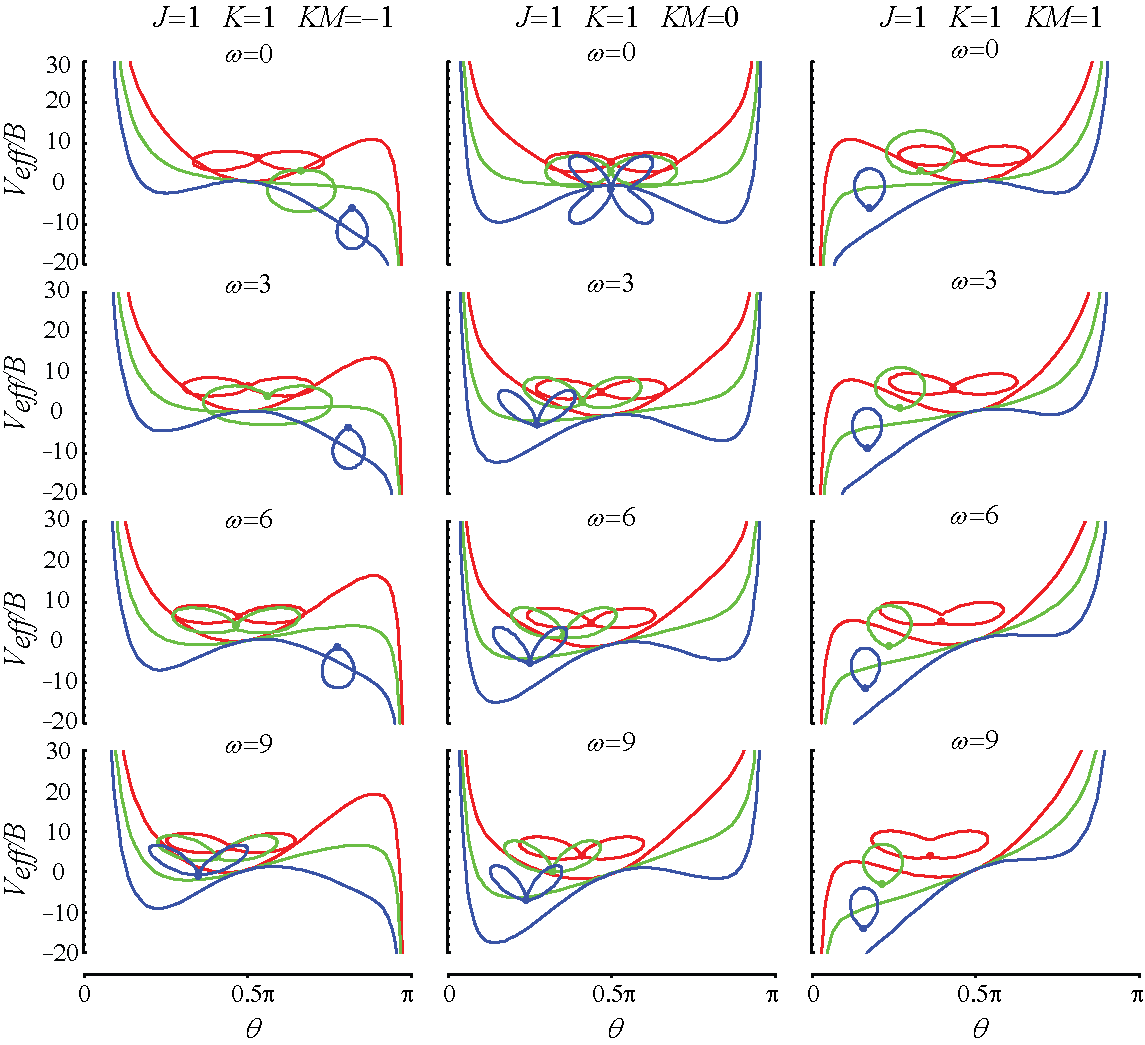}
\caption{(color online) Effective potential, $U$, for $K=1$ and $MK=-1$ (left panels), $MK=0$ (center panels), and $MK=1$ (right panels) along with the eigenenergies and orientation amplitudes (shown by dots) and squares of the wavefunctions for states with $\tilde{J}=1$. The column are comprised of panels pertaining to increasing values of $\omega$. Red curves correspond to $\Delta\omega =-15$, green curves to $\Delta \omega =0$, and blue curves to $\Delta\omega =15$. See text. \label{fig:EffPotOmega1}}
\end{figure*}

\begin{figure*}[htbp]
\centering
\includegraphics[]{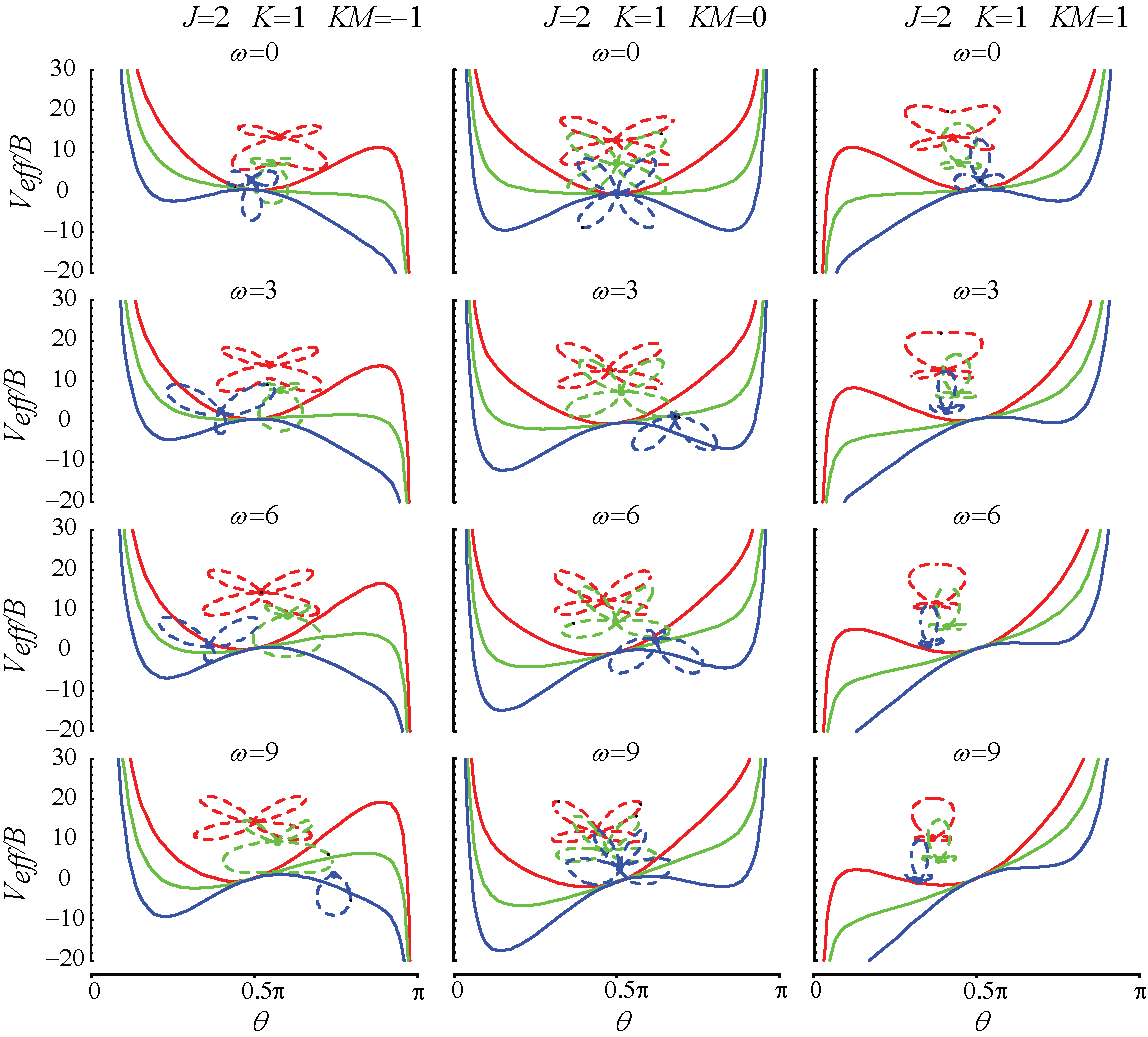}
\caption{(color online) Effective potential, $U$, for $K=1$ and $MK=-1$ (left panels), $MK=0$ (center panels), and $MK=1$ (right panels) along with the eigenenergies and orientation amplitudes(shown by dots) and squares of the wavefunctions for states with $\tilde{J}=2$. The column are comprised of panels pertaining to increasing values of $\omega$. Red curves correspond to $\Delta \omega =-15$, green curves to $\Delta \omega =0$, and blue curves to $\Delta\omega =15$. See text.\label{fig:EffPotOmega2}}
\end{figure*}

The molecular-axis orientation by which the synergism of the static and radiative fields for $\boldsymbol{\alpha}$ oblate manifests itself can be best seen in Fig. \ref{fig:Electric}f and interpreted with the help of the effective potential, Figures  \ref{fig:EffPotOmega1} and \ref{fig:EffPotOmega2}. Figs. \ref{fig:Laser}d,e,f provide additional cuts through the same $(\omega ,\Delta \omega )$ parameter space. Conspicuously, all states, for $\boldsymbol{\alpha}$ oblate, become right-way oriented at a sufficiently large $\omega $, cf. Fig. \ref{fig:Electric}f. However, some of the states either become (e.g., $|2,1,0\rangle $) or are (e.g., $|1,1,-1\rangle $) wrong-way oriented first. As $\omega $ increases, the $|2,1,-1\rangle $ state is even seen to become right-way oriented, then wrong-way oriented, and finally right way oriented again. This behavior is a consequence of the different types of coupling that the states in question are subjected to. We'll discuss them in turn.

The $|2,1,0\rangle $ state is the upper member of a tunneling doublet whose lower member is the $|1,1,0\rangle $ state, cf. Fig. \ref{fig:CorrelationLaser}. At $\omega =0$, the $|1,1,0\rangle $ and $|2,1,0\rangle $ states are not oriented, as is the case for any states with $KM=0$. However, the value of $\Delta \omega =15$ is large enough to push the levels into quasi-degeneracy, see Fig. \ref{fig:Laser}a, in which case the static field can easily couple them. But at $\omega \ll \Delta \omega $, such a coupling results in localizing the wavefunctions of the $|1,1,0\rangle $ and $|2,1,0\rangle $ pair in the forward and backward wells, respectively, of the effective potential $U$, which, for $\omega \ll \Delta \omega $, are mainly due to the polarizability interaction. As $\omega $ becomes comparable to $\Delta \omega $, the effective potential becomes skewed. The forward well grows deeper at the expense of the backward well and the wrong-way oriented $|2,1,0\rangle$ state is flushed out into the forward well as a result, thus acquiring the right-way orientation. The blue effective potentials and wavefunctions in the middle panels of Figs. \ref{fig:EffPotOmega1} and \ref{fig:EffPotOmega2} detail this behavior. We note that the lower and upper member of a given tunneling doublet is always right- and wrong-way oriented, respectively, at $\omega \ll \Delta \omega $. This is because the coupling by $V_{\mu }$ makes the states to repel each other, whereby the upper level is pushed upward and the lower level downward. The noted orientation of the two states then immediately follows from the Hellmann-Feynman theorem.

The $|1,1,-1\rangle $ state has $\tilde{J}<|M|+|K|$ and thus, in the radiative field alone, is a member of a degenerate doublet, along with the $ |1,1,1\rangle $ state, cf. the blue effective potentials and wavefunctions in the upper left and right panels of Fig. \ref{fig:EffPotOmega1}. While the $|1,1,1\rangle $ state is always right way oriented, the $|1,1,-1\rangle $ state is wrong-way oriented even at $\omega =0$ thanks to the asymmetry of the effective potential due to its angle-dependent third term, proportional to $KM$, cf. eq. (\ref{eqn:effpot}). An increase in $\omega $ removes the degeneracy of the doublet and causes the wrong-way oriented $|1,1,-1\rangle $ state to have a higher energy than the right-way oriented $|1,1,1\rangle$ state. As $\omega $ increases, $V_{\mu }$ deepens the forward well and, as a result, the wavefunction of the $|1,1,-1\rangle$ state rolls over into it, thus making the state right-way oriented.

The $|2,1,-1\rangle $ state exhibits an even more intricate behavior. Instead of a wrong-way orientation at low $\omega$, enhanced by the radiative field, the state becomes right way oriented first, due to an avoided crossing with the $|1,1,-1\rangle $ state (whose behavior, sketched above, is, of course, also affected by the same avoided crossing). This is followed, at increasing $\omega $, by a ``native'' wrong-way orientation that, at $\omega \gtrsim 20$, is reversed by virtue of the deepening forward well of the effective potential, which confines the state.

We note that the tangle of level crossings seen in Figs. \ref{fig:Electric} and \ref{fig:Laser} that complicate the directional properties of symmetric tops in the combined fields is caused by the reversed ordering of the energy levels due to the permanent and induced dipole interactions: as the static and radiative fields are cranked up, the levels ``comb'' through one another.

\subsection{Perpendicular fields\label{subsec:PerpendicularFields}}

For a tilt angle $\beta \neq 0$ or $\pi $ between the static, $\varepsilon _{S}$, and radiative, $\varepsilon _{L}$, fields, the combined-fields problem loses its cylindrical symmetry and $M$ ceases to be a good quantum number. This greatly contributes to the complexity of the energy level structure and the directional properties of the states produced. At the same time, the $M$-dependent effective potential, eq. (\ref{eqn:effpot}), so useful for understanding the directionality of the states produced by the collinear fields, cannot be applied to the case of perpendicular fields, as $M$ is not defined.

\begin{figure*}[htbp]
\centering
\includegraphics[width=\textwidth]{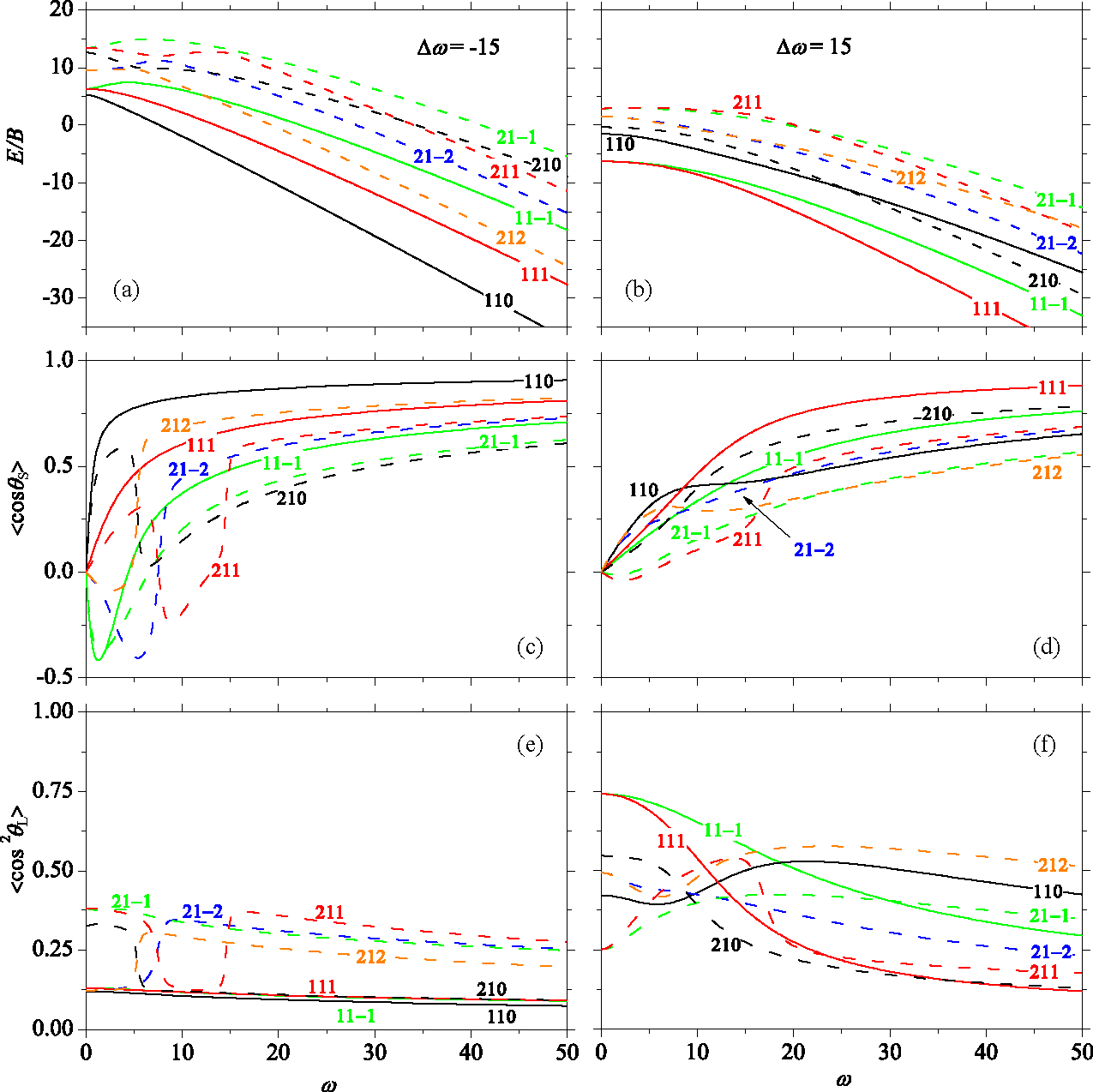}
\caption{(color online) Dependence, in perpendicular fields, of the eigenenergies, panels (a)-(b), orientation cosines, panels(c)-(d), and alignment cosines, panels (e)-(f), of the states with $0<\tilde{J}\leq 3$, $-1\leq MK\leq 1$, and $K=1$ on the dimensionless parameter $\omega$ (that characterizes the permanent dipole interaction with the electrostatic field) for fixed values of the parameter $\Delta\omega $ (that characterizes the induced-dipole interaction with the radiative field; $\Delta \omega <0$ for prolate polarizability anisotropy, $\Delta \omega >0$ for oblate polarizability anisotropy). The states are labeled by $|\tilde{J}KM\rangle$. The orientation cosines $\langle \cos\theta _{S}\rangle$ are calculated with respect to the electrostatic field and the alignment cosines $\langle \cos^{2}\theta _{L}\rangle$ with respect to the laser field. See text. \label{fig:PF-Electric}}
\end{figure*}

\begin{figure*}[htbp]
\centering
\includegraphics[width=\textwidth]{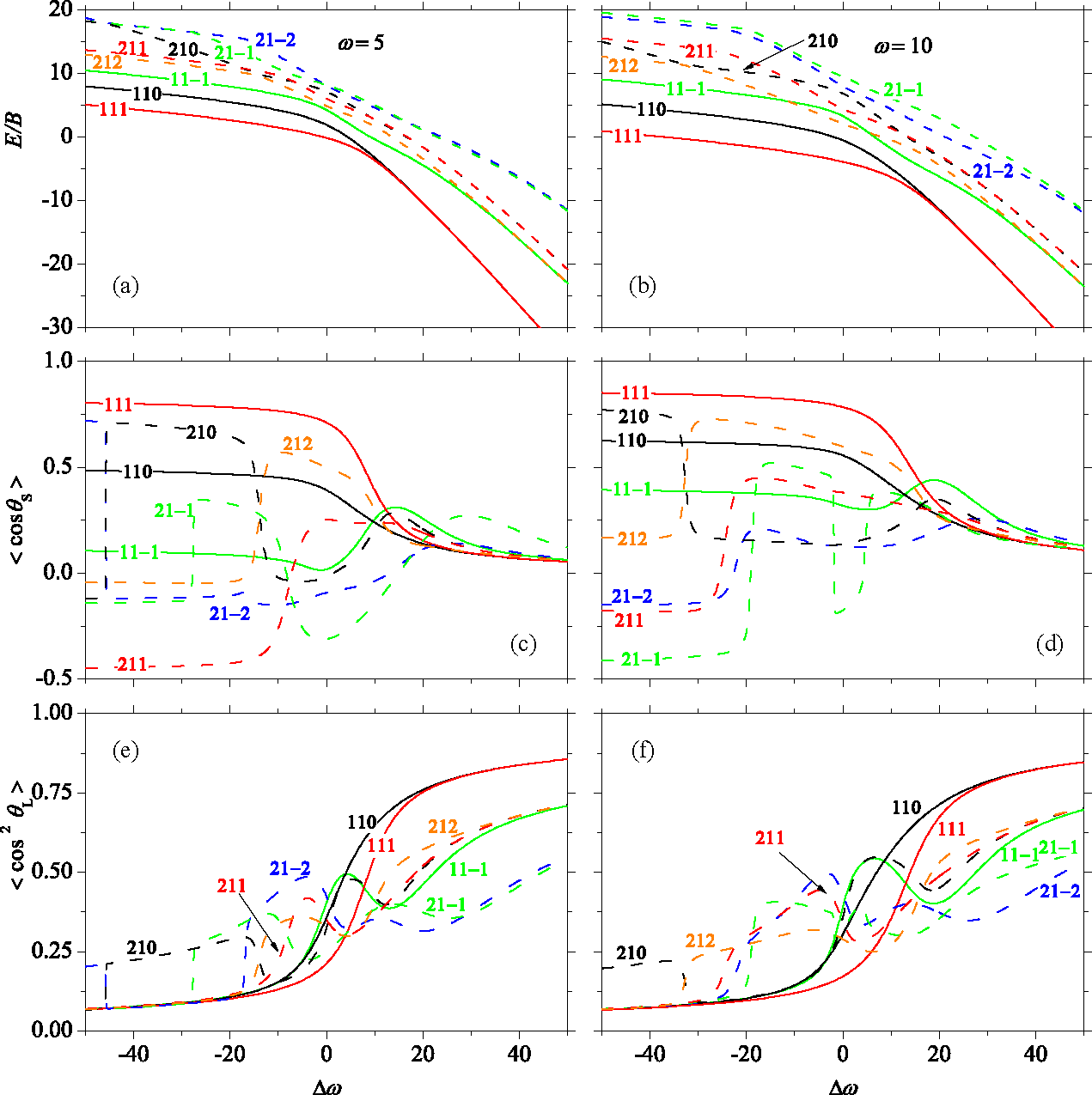}
\caption{(color online) Dependence, in perpendicular fields, of the eigenenergies, panels (a)-(b), orientation cosines, panels(c)-(d), and alignment cosines, panels (e)-(f), of the states with $0<\tilde{J}\leq 3$, $-1\leq MK\leq 1$, and $K=1$ on the dimensionless parameter $\Delta \omega $ (that characterizes the induced-dipole interaction with the radiative field; $\Delta \omega <0$ for prolate polarizability anisotropy, $\Delta \omega >0$ for oblate polarizability anisotropy) for fixed values of the parameter $\omega $ (that characterizes the permanent-dipole interaction with the electrostatic field). The states are labeled by $|\tilde{J}KM\rangle $. The orientation cosines $\langle \cos \theta _{S}\rangle $ are calculated with respect to the electrostatic field and the alignment cosines $\langle \cos^{2}\theta _{L}\rangle $ with respect to the laser field. See text.\label{fig:PF-Laser}}
\end{figure*}

Figures \ref{fig:PF-Electric} and \ref{fig:PF-Laser} show the dependence of the eigenenergies and of the orientation and alignment cosines on the field strength parameters $\omega $ and $\Delta \omega $ for a similar set of states as in Figs. \ref{fig:Electric} and \ref{fig:PF-Laser} for the collinear fields.

Figs. \ref{fig:PF-Laser}c and d capture well the main patterns of the behavior. For $\boldsymbol{\alpha }$ prolate, the interaction with the radiative field $\varepsilon _{L}$ aligns the body-fixed electric dipole $\boldsymbol{\mu} $ along the perpendicular static field $\varepsilon _{S}$, cf. Fig. \ref{fig:FieldsMolecules}. For $\boldsymbol{\alpha }$ oblate, $\varepsilon _{L}$ aligns $\boldsymbol{\mu}$ perpendicular to $\varepsilon _{S}$. As a result, in perpendicular fields, $\boldsymbol{\alpha}$ prolate yields a strong orientation whereas $\boldsymbol{\alpha}$ oblate a vanishing one. This is the inverse of the situation in collinear fields. However, since $V_{\alpha}$ prolate is a single-well potential, the levels lack the patterns found for collinear fields for $ V_{\alpha }$ oblate. Due to the multitude of avoided crossings, the states often switch between the right and wrong-way orientation, even over tiny ranges of the interaction parameters. Therefore, a much finer control of the parameters is needed in the case of perpendicular fields in order to preordain a certain orientation. For $\boldsymbol{\alpha }$ oblate, the coupling of the different states is weak, and the avoided crossings that abound in the parallel case, Fig. \ref{fig:Laser}, are almost absent, see Fig. \ref{fig:PF-Laser}.

The states are essentially all high-field seeking in the radiative field for $\Delta \omega >0$ and low-field seeking for $\Delta \omega <0$. This reflects the repulsive and attractive character of the polarizability interaction in the prolate and oblate case, respectively. In the oblate case, the states shown are essentially high-field seeking in the static field. This means that for $\Delta \omega >0$, much of the wrong-way orientation seen, e.g., in Fig. \ref{fig:Electric}d,f, can be eliminated, see Fig. \ref{fig:PF-Electric}d. Unfortunately, the elimination of the wrong-way orientation happens at the expense of the magnitude of the orientation, which remains small.

Depending on the relative strength of the two fields, the effects of one can dominate those of the other. This contrasts with the behavior in the collinear fields where even a tiny admixture of the static field can dramatically change the behavior of the states due to the radiative field (such as the coupling of the tunneling doublets).

A detailed comparison of the effect the two fields have on a given state is complicated by the dependence of the state label on the sequence in which the parameters $\omega $, $\Delta\omega$ and $\beta$ are varied. This behavior is further analyzed below.

As the dependence of the orientation cosine on the $\omega $ parameter indicates, see Fig. \ref{fig:PF-Electric}c, the largest orientation for a prolate polarizability is attained for the $\left| 1,1,0\right\rangle $ state. Other states, such as the $\left|1,1,-1\right\rangle $ state, become right-way oriented only for sufficiently large field strengths. At $\omega $ large, the orientation of all states becomes substantially greater than what is achievable with collinear fields, cf. Fig. \ref{fig:Electric}e. For the oblate polarizability, the states behave similarly, exhibiting a uniform behavior. On the other hand, in dependence on $\Delta \omega $, Fig. \ref{fig:PF-Laser}, the $\left| 1,1,1\right\rangle $ state - instead of the $ \left|1,1,0\right\rangle $ state - exhibits the strongest right-way orientation over most of the range shown. This is an example of the sequence dependence of the state label. The $\left| 2,1,-1\right\rangle $ state shows the most dramatic variations with $\Delta\omega $. It has three rather sharp turn-around points, where the direction of the dipole moment changes. Another example of the sequence dependence of the label is the $ \left| 2,1,1\right\rangle $ state in Fig.\ref{fig:PF-Electric}c, which is wrong-way oriented at $\omega \approx 10$, whereas none of the states shown in Figure \ref{fig:PF-Laser}d is wrong-way oriented for $\Delta \omega$ ranging from $-20$ to $0$.

When $\Delta\omega\gg\omega$, quasi-degenerate states are formed, similar to the tunneling doublets. However, the electrostatic field is not able to couple them as a result of which the states are only aligned.

\begin{figure*}[htbp]
\centering
\includegraphics[]{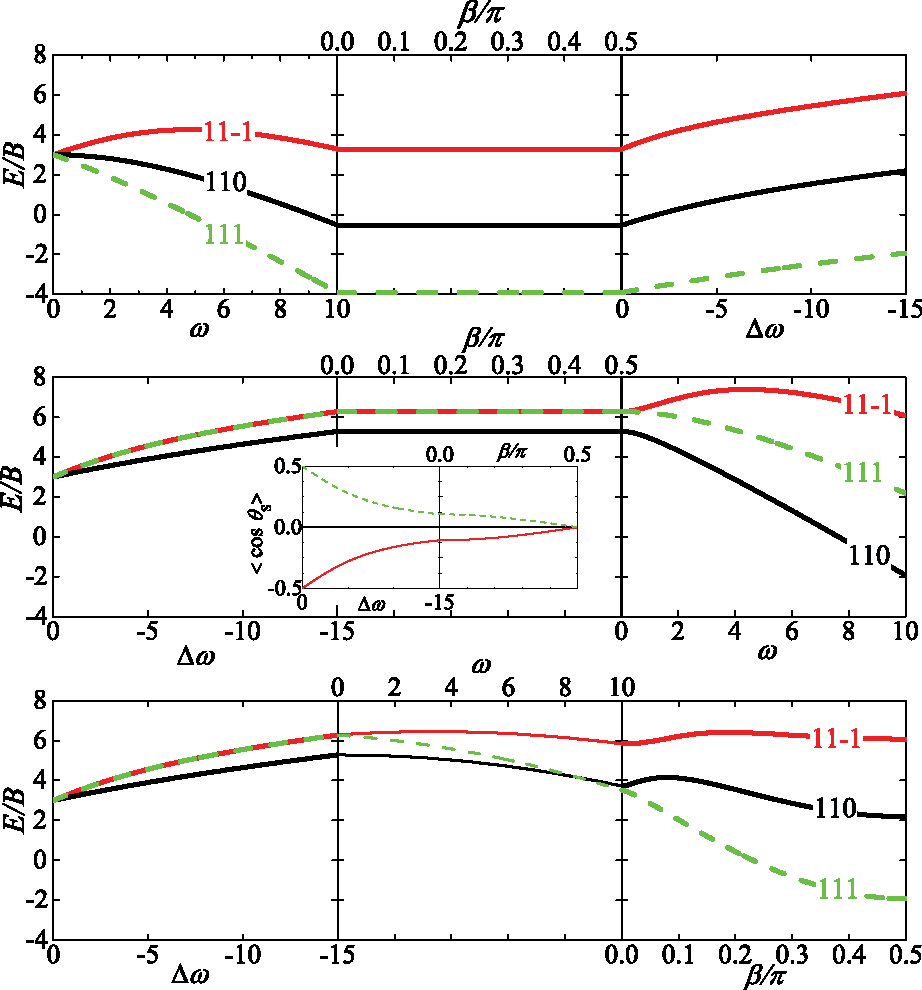}\caption{
(color online) Comparison of the evolution of states for different path through the parameter space. From top to bottom these are: $\left|\tilde{J}K\tilde{M}; \omega, \beta, \Delta\omega\right\rangle$, $\left|\tilde{J}K\tilde{M}; \Delta\omega, \beta, \omega\right\rangle$ and $\left| \tilde{J}K \tilde{M}; \Delta\omega, \omega, \beta\right\rangle$. In the middle panel the $\tilde{1}1\tilde{0}$ state is interchanged with the $\tilde{1}1\tilde{-1}$ state in the other two panels. We dub this effect `label switching.'\label{fig:PfadVgl}}
\end{figure*}

In tilted fields, the label of a given state may depend on the sequence in which the fields are switched on and the tilt angle spanned by the field vectors is varied. Figure \ref{fig:PfadVgl} shows the adiabatic evolution of the states with $\tilde{J}=1$, $K=1$ and $\tilde{M}=-1,0,1$ for three different sequences leading to a crossing points in Figures \ref{fig:PF-Electric} and \ref{fig:PF-Laser} at $\omega =10$, $\Delta \omega =-15$ and $\beta =\pi /2$ . The three sequences are: $(1)$ The electrostatic field is turned on to a value such that $\omega =10$; then the laser field is switched on to a value such that $\Delta \omega=-5\cdot 10^{-5}$; then the tilting of the fields is carried out to $\beta =\pi /2$; finally $\Delta \omega $ is raised in steps of $5\cdot 10^{-5}$ up to $\Delta \omega =-15$. The corresponding states are labeled as $|\tilde{J},K,\tilde{M};\omega ,\beta ,\Delta \omega \rangle $. $(2)$ The laser field is turned up to a value such that $\Delta \omega =-15$; then the electrostatic field is switched on to a value such that $\omega =5\cdot 10^{-5}$; the fields are tilted to $\beta =\pi /2$; $\omega $ is raised (in steps of $5\cdot 10^{-5})$ to $\omega =10$. The states are labeled as $|\tilde{J},K,\tilde{M};\Delta \omega ,\beta ,\omega \rangle$. $(3)$ The laser field is turned up to a value such that $\Delta\omega$. Subsequently, the electrostatic field is turned on to a value such that $\omega =10$; finally, the fields are tilted to $\beta =\pi /2$. The corresponding states are labeled by $|\tilde{J},K,\tilde{M};\Delta \omega ,\omega ,\beta \rangle .$ We note that the states $|\tilde{J},K,\tilde{M};\omega ,\Delta \omega,\beta \rangle $ yield identical results with those obtained for sequence (3).

As one can see in Fig. \ref{fig:PfadVgl}, there are no discontinuities for any of the calculations that might suggest that our state-tracking algorithm has failed. However, the three sequences yield different labels for the same state at the crossing point. The $|1,1,0;\Delta \omega ,\beta ,\omega \rangle$ and the $|1,1,1; \omega ,\beta ,\Delta\omega \rangle $ states change their labels for sequence 2 as compared with sequences 1 and 3. We note that the absence of genuine crossings for sequence 1 and 2 precludes the possibility that the tracking algorithm jumped an avoided crossing. In the middle panel of sequence 2, the $M=1$ and $M=-1$ states appear degenerate, but they are not. The very weak electric field, necessary to give a meaning to the mutual tilting of the fields, introduces a small splitting. The wavefunctions and thus the orientation is different for these two states. The orientation cosines for all three states of the middle panel for sequence 2 are shown in the panel's inset.

To our knowledge the above phenomenon of \emph{label switching} has not been previously described. We find that it occurs not only for states with the same $\tilde{J}$ and different $\tilde{M}$ but also for states with different $\tilde{J}$.

Two mechanisms seem likely to be responsible for label switching: (i) symmetry breaking; (ii) chaotic behavior of the underlying
classical system and the concomitant singularities of the classical phase space.

The first mechanism is at hand whenever several interactions are present, only some of which break the symmetry of the system completely. Whereas the symmetry-conserving interactions give rise to a block structure of the Hamiltonian matrix and so allow for genuine crossings of states belonging to different blocks, the symmetry breaking interaction leads to just a single block and makes all crossings of the states avoided. If the symmetry-conserving interactions are turned on first, the system may have undergone a genuine crossing before the symmetry-breaking interaction is turned on second. However, the opposite is not true, as the symmetry-breaking interaction, when turned on first, makes all crossings avoided once and for all.

However, in the case of our system, there seems to be a symmetry left, which prevents the system from possessing avoided crossings only. For sufficiently high densities of points in the parameter space, one would expect to make any avoided crossings, if present, visible. Nevertheless, even at the highest point densities, many of the crossings have been found to be genuine.  For a closed path in the parameter space at $\beta \neq 0$, the evolution of states exhibits label switching as well. This is the case for closed paths defined in terms of $\omega $ \& $\beta $ (at fixed $\Delta \omega $) as well as in terms of $\Delta \omega $ \& $\omega $ (at fixed $\beta $ - in which case the symmetry does not change). Closed paths defined in terms of $ \Delta \omega $ \& $\beta $ (at fixed $\omega $) do not lead to label switching, which suggests that the permanent dipole interaction plays a major role. From this we infer that label switching is not only due to symmetry breaking. We checked whether the linear rotor in the combined tilted fields also exhibits label switching, and found that it does.

Arango \emph{et al.} \cite{Arango2004} and Kozin and Roberts \cite{Kozin2003} have shown that a diatomic and a symmetric top in an electrostatic field alone exhibit classical and quantum \emph{monodromy }\cite{Sadovskii2006}. In quantum mechanics, monodromy reveals itself as a defect in the discrete lattice of states.

For tilted fields, $M$ ceases to be a good quantum number, and its classical analog, $m$, is no longer a constant of motion for a classical diatomic or a symmetric top in tilted fields. As a result, the problem becomes nonintegrable. For tilt angles $\beta \sim \pi /4$, Arango \emph{et al.} found that the diatomic exhibits extensive classical chaos \cite{Arango2004}, \cite{Arango05}, while Kennerly mentions quantum chaos for symmetric tops \cite{PhdKennerly05}. In order to see monodromy for the tilted fields problem, one would wish to construct a quantum lattice. However, it is not clear how to go about it in the absence of good quantum numbers (apart from $K$). A possibility is to make use of the root-mean-square expectation value of $M$, $\langle M^{2}\rangle ^{1/2}$, and construct a quasi-monodromy diagram with its aid. Such a diagram can be then compared with classical results, since $\langle m^{2}\rangle ^{1/2}$ is well-defined in classical calculations \cite{Arango2004}. The quasi-monodromy diagrams that we so constructed strongly depend on the ratio of the two field strength parameters. The structure of the diagrams for the two types of polarizability anisotropy differs markedly, although it's the electrostatic field alone that is supposed to determine the behavior. 
The quasi-monodromy diagrams are further complicated by the dependence on the representation of the angular variables, as $\left\langle M^2\right\rangle^{1/2}$  differs in the two representations. Only $\left\langle M\right\rangle$ is well-defined and independent of the representation. The quasi-monodromy diagrams can be obtain on request.

Hence, we found that in titled fields switching of the adiabatic labels of states for linear molecules as well as symmetric tops takes place, where it arises in connection with the change of the character of a crossing, from genuine to avoided. Monodromy, however, is present for both collinear and tilted fields. We note that whether label switching could be experimentally observed and utilized is not clear at this point.

\section{Examples and applications\label{Appl}}

Table \ref{tbl:examples} lists a swatch of molecules that fall under the various symmetry combinations of the $\boldsymbol{\alpha }$ and $\mathbf{I}$\ tensors, as defined in Table \ref{tbl:SymComb}. The table lists the rotational constants, dipole moments, polarizability anisotropies as well as the values of the interaction parameters $\omega $ and $\Delta \omega $ attained, respectively, at a static field strength of $1$ kV/cm and a laser intensity of $10^{12}$ W/cm$^{2}$. The conversion factors are also included in the table. While the field strength of the electrostatic field of a kV/cm is easy to obtain (or sometimes even difficult to avoid), a laser intensity of a petawatt per cm$ ^{2}$ is somewhat harder to come by. However, pulsed laser radiation can be easily focused to attain such an intensity, and a nanosecond pulse duration is generally sufficient to ensure adiabaticity of the hybridization process.

\begin{table}[h]
\centering
\begin{tabular}{c|c|c|c|c|c}
\hline\hline Molecule \T  & $B$  & $\mu$ & $\Delta \alpha $
 & $\omega$ & $\Delta \omega $ \\
 & [MHz] & $[D]$ & [\r{A}$^{3}$] & @$1$  & @$10^{12}$ \\ 
 \B&  &	 &  &  kV/cm &  W/cm$^{2}$  \\ \hline
Acetonitrile \T & $9199$ & $3.92$ & $1.89$ & $0.22$ & $65.0$ \\
Ammonia & $298500$ & $1.47$ & $0.24$ & $0.0025$ & $0.24$ \\
Benzene-Ar & $1113$ & $(0.1)$ & $-6.1$ & $(0.05)$ & $-1735$ \\
Bromomethane & $9568$ & $1.82$ & $1.95$ & $0.10$ & $64.5$ \\
Chloromethane & $13293$ & $1.89$ & $1.69$ & $0.07$ & $40.2$ \\
Fluoromethane & $10349$ & $1.85$ & $0.84$ & $0.09$ & $25.7$ \\
Iodomethane & $7501$ & $1.64$ & $2.15$ & $0.11$ & $90.7$ \\
Trichlormethane & $3302$ & $1.04$ & $-2.68$ & $0.16$ & $-257$ \\
Trifluoromethane \B & $7501$ & $1.65$ & $-0.18$ & $0.11$ & $-7.60$ \\
\hline\hline
\end{tabular}
\caption{\label{tbl:examples} Values of parameters $\omega $ and
$\Delta \omega $ for choice symmetric top molecules whose
properties were taken from \cite{Pogliani2003,Beilstein}. The conversion factors are: $\omega =503.2$ $\mu $ $[D]$ $\varepsilon _{S}$ [kV/cm]$/B$ [MHz] and $\Delta \omega=3.1658\times 10^{-7}$ $I$ [W/cm$^{2}$] $\Delta \alpha $
[\r{A}$^{3}$]$/B$ [MHz]$.$ Numbers in parentheses are order-of-magnitude estimates.}
\end{table}

We note that the directional properties displayed in Figs. \ref{fig:Electric} -\ref{fig:PF-Laser} should be attainable for most of the molecules listed in Table \ref{tbl:examples}.

The amplification of molecular orientation in the combined fields may find a number of new applications.

In molecule optics \cite{Zhao2003}, a combination of a pulsed nonresonant radiative field with an inhomogeneous electrostatic field can be expected to give rise to temporally controlled, state-specific deflections. In an inhomogeneous static field produced by an electrode micro-array \cite{Schulz2004}, and a ns laser pulse of $10^{12}$ W/cm$^{2}$, deflections on the order of a mrad appear feasible for light molecules. Since the synergistic effect of the combined static and radiative fields is only in place when both fields are on, the deflection of the molecules would be triggered by the presence of the ns laser pulse. The sensitivity of the deflection process to the magnitude of the space-fixed electric dipole moment and its direction (right- or wrong-way) would simultaneously enable state-selection.

The directional properties of symmetric tops in the combined fields may come handy in the studies of rare-gas molecules of the type found by Buck and Farnik. Some of the species identified by these authors may in fact possess a three-fold or higher axis of rotation symmetry \cite{Buck2006}, \cite{farnik}.

Drawing on the analogy with previous work on spectral effects in single electrostatic \cite{ROST1992}, \cite{Wu1994} or radiative fields \cite{FRIEDRICH1995}, we also note that the combined fields can be expected to dramatically modify the spectra of symmetric top molecules: this is due to a change of both energy levels and the transition dipole moments. The latter are strongly affected by the directionality of the wavefunctions, which give rise to their widely varying overlaps. However, detailed simulations of such effects in the combined fields are still wanting.

Last but not least, molecules in tilted electrostatic and radiative fields can serve as prototypical systems for the study of quantum chaos. The absence of good quantum numbers and the multitude of unstable equilibria suggest this possibility. Indeed, as pointed out in the work of the Ezra group \cite{Arango2004}, \cite{Arango05}, linear molecules in non-collinear fields exhibit chaotic dynamics. We hope that our present study of symmetric tops in the combined fields will provide an impetus for further work on quantum chaos and monodromy exhibited by such nonintegrable molecular systems.

\section{Conclusions\label{sec:Concl}}

In our theoretical study of the directional properties of symmetric top molecules in combined electrostatic and nonresonant radiative fields, we saw that collinear (perpendicular) fields force permanent dipoles of molecules with oblate (prolate) polarizability anisotropy into alignment with the static field. We found that the amplification mechanism that produces highly oriented states for linear molecules ($K=0$) is also in place for precessing states of symmetric tops with $\tilde{J}\geq |K|+|M|$. This mechanism is based on the coupling by a collinear electrostatic field of the tunneling doublet states created by the interaction of the oblate polarizability with a linearly polarized radiative field. The efficacy of this coupling is enhanced by an increased strength, $\Delta \omega >0$, of the anisotropic polarizability interaction that traps the tunneling doublets it creates deeper in a double-well potential, and draws the members of the doublets closer to one another. Apart from this synergistic effect of the combined fields, there is another effect in place, but for states with $\tilde{J} <|K|+|M| $. Such states occur as exactly degenerate doublets in the radiative field alone, whose both members are strongly but mutually oppositely oriented along the polarization plane of the field. This orientation can be manipulated by adding an electrostatic field, which easily couples the right-way oriented states to its direction, whether this is parallel or perpendicular to the radiative field. At a sufficiently large strength, $\omega $, of the permanent dipole interaction, the wrong-way oriented member of the (no longer) degenerate doublet becomes also right-way oriented. The above patterns of the energy levels are complicated by numerous avoided crossings, themselves unavoidable, due to the opposite ordering of the energy levels for the permanent and induced dipole interaction.

The absence of cylindrical symmetry for perpendicular fields is found to preclude the wrong-way orientation for the oblate polarizability, causing all states to become high-field seeking with respect to the static field. The changes of the system's parameters $\omega $, $\Delta \omega $, and $ \beta $ cause genuine crossings to become avoided and vice versa. This, in turn, causes the eigenstates to follow different adiabatic paths through the parameter space and to end up with adiabatic labels that depend on the paths taken.

The amplification of molecular orientation by the synergistic action of the combined fields may prove useful in molecule optics and in spectroscopy. Monodromy and quantum chaos lurk behind the combined fields effects whose further analysis may thus be expected to shed new light on both.\\

{\bf Acknowledgments}\\

We thank Dmitrii Sadovskii, B. Zhilinskii, and Igor Kozin for discussions related to the issue of label switching. We are grateful to Gerard Meijer for fruitful discussions and support.\\

\begin{appendix}
\section{Appendix\label{MatrixElements}}
It is expedient to represent the elements of the Hamiltonian matrix in the basis of the symmetric-top wavefunctions, $\left|
JKM\right\rangle$. These are related to the Wigner functions $D^J_{MK}$ via
\begin{eqnarray}
\left|JKM\right\rangle\!=\!(-\!1)^{M\!-\!K}\!\left[\frac{2J+1}{8
\pi^2}\right]^{1/2}\!D^J_{-\!M-\!K}\left(\varphi,\theta,\chi\right)
\end{eqnarray}

The field operators can be expressed in terms of the Legendre polynomials $P_J(\cos\theta)$ and spherical harmonics $Y_{JM}(\theta,\varphi)$, which in turn, are related to the Wigner functions by
\begin{eqnarray}
D^J_{M0}\left(\varphi,\theta,\xi\right)&=&\left(\frac{4\pi}{2J+1}\right)^{1/2}Y^*_{JM}\left(\theta,\varphi\right)\\
D^J_{00}\left(\varphi,\theta,\xi\right)&=&P_J\left(\cos\theta\right)
\end{eqnarray}

The evaluation of the matrix elements of the Hamiltonian then only requires the application of the ``triple product over
Wigner functions" theorem,
\begin{align*}
\int D^{J_3}_{M_3 K_3}(\textbf{R})D^{J_2}_{M_2 K_2} & (\textbf{R})D^{J_1}_{M_1
K_1}(\textbf{R})d\Omega=\nonumber\\
& 8\pi^2 \left(\begin{array}{ccc}
J_1 &   J_2 &   J_3\\
M_1 &   M_2 &   M_3
\end{array}\right)
\left(\begin{array}{ccc}
J_1 &   J_2 &   J_3\\
K_1 &   K_2 &   K_3
\end{array}\right)
\end{align*}
where $\textbf{R}$ denotes $\left(\varphi,\theta,\xi\right)$.

(A) For the collinear case, the matrix elements to be determined are:
\begin{eqnarray}
\left\langle
J^{\prime}K^{\prime}M^{\prime}\left|\frac{H}{B}\right|\right.&& JKM\bigg\rangle=\nonumber\\
&&\left\langle J^{\prime}K^{\prime}M^{\prime}\left|J(J+1)+\rho
K^2- \omega_\perp \right|JKM\right\rangle \nonumber\\&&-
\omega\left\langle
J^{\prime}K^{\prime}M^{\prime}\left|\cos\theta\right|JKM\right\rangle
\nonumber\\&&- \Delta\omega\left\langle
J^{\prime}K^{\prime}M^{\prime}\left|\cos^2\theta\right|JKM\right\rangle
\end{eqnarray}
where
\begin{eqnarray}
\cos\theta&=&P_1(\cos\theta)=D^1_{00}(\varphi,\theta,\xi)\\
\cos^2\theta&=&\frac{1}{3}\!\left(2
P_2(\cos\theta)\!+\!1\right)\!=\!\frac{1}{3}\!\left(2
D^2_{00}(\varphi,\theta,\xi)\!+\!1\right)
\end{eqnarray}
The properties of the 3j-symbols eliminate interactions between states with different $M$ and $K$. As a result, states belonging
to certain $M$ and $K$ values can be treated in a separate calculation.
\begin{align*}
\bigg\langle J^{\prime}KM & \left. \left|\frac{H}{B}  \right|JKM\right\rangle=\delta_{JJ^{\prime}}\left(J(J+1)+\rho K^2 -\omega_{\perp}\right)\\
&-\omega (2J+1)^{1/2}(2J^{\prime}+1)^{1/2}\nonumber\\&\times(-1)^{M-K}
\underbrace{\left(
\begin{array}{ccc}
J   &   1   &   J^{\prime}\\
-M  &   0   &   M
\end{array}\right)}_{J^{\prime}=J,J\pm1}
\underbrace{\left(
\begin{array}{ccc}
J   &   1   &   J^{\prime}\\
-K  &   0   &   K
\end{array}\right)}_{J^{\prime}=J,J\pm1}\\
&-\Delta\omega
\frac{2}{3}(2J+1)^{1/2}(2J^{\prime}+1)^{1/2}\nonumber\\&\times(-1)^{M-K}\underbrace{\left(
\begin{array}{ccc}
J   &   2   &   J^{\prime}\\
-M  &   0   &   M
\end{array}\right)}_{J^{\prime}=J,J\pm1,J\pm2}
\underbrace{\left(
\begin{array}{ccc}
J   &   2   &   J^{\prime}\\
-K  &   0   &   K
\end{array}\right)}_{J^{\prime}=J,J\pm,J\pm2}
\end{align*}

(B) For the case of tilted fields, the matrix elements can be expressed either in terms of the static field or the laser
field. The `simpler' case is keeping the laser field fixed in space and tilting the electric field relative to it. Then the
$\cos\theta$ operator is replaced by:
\begin{eqnarray}
\cos\theta_s&=&\cos\beta\cos\theta+\sin\beta\sin\theta\cos\varphi\nonumber\\
&=&\cos\beta
D^1_{00}+\sin\beta\sqrt{\frac{1}{2}}\left(D^1_{-10}-D^1_{10}\right)\nonumber
\end{eqnarray}
Different $K$ states do not mix. The matrix elements are:
\begin{align*}
&\left\langle\! J^{\prime}KM^{\prime}\!\left|\!\frac{H}{B}\!\right|\!JKM\!\right\rangle\!=\!\delta_{J\!J^{\prime}}\delta_{M\!M^{\prime}}\left(J(J\!+\!1)\!+\!\rho K^2 \!-\!\omega_{\perp}\right)\\
&-\omega \cos\beta(2J+1)^{1/2}(2J^{\prime}+1)^{1/2}\nonumber\\&\hspace*{1cm}(-1)^{M-K}
\underbrace{\left(
\begin{array}{ccc}
J   &   1   &   J^{\prime}\\
-M  &   0   &   M
\end{array}\right)}_{J^{\prime}=J,J\pm1}
\underbrace{\left(
\begin{array}{ccc}
J   &   1   &   J^{\prime}\\
-K  &   0   &   K
\end{array}\right)}_{J^{\prime}=J,J\pm1}\\
&-\omega (2J+1)^{1/2}(2J^{\prime}+1)^{1/2}(-1)^{M-K}
\sin\beta\sqrt{\frac{1}{2}}\\
&\hspace*{1cm}\times\Bigg[ \underbrace{\left(
\begin{array}{ccc}
J   &   1   &   J^{\prime}\\
-M  &   -1   &   M^{\prime}
\end{array}\right)}_{J^{\prime}=J,J\pm1;M^{\prime}=M+1}
\underbrace{\left(
\begin{array}{ccc}
J   &   1   &   J^{\prime}\\
-K  &   0   &   K
\end{array}\right)}_{J^{\prime}=J,J\pm1}\nonumber\\&\hspace*{1.5cm}-
\underbrace{\left(
\begin{array}{ccc}
J   &   1   &   J^{\prime}\\
-M  &   1   &   M^{\prime}
\end{array}\right)}_{J^{\prime}=J,J\pm1;M^{\prime}=M-1}
\underbrace{\left(
\begin{array}{ccc}
J   &   1   &   J^{\prime}\\
-K  &   0   &   K
\end{array}\right)}_{J^{\prime}=J,J\pm1}\Bigg] \\
&-\Delta\omega
(2J+1)^{1/2}(2J^{\prime}+1)^{1/2}\nonumber\\&\hspace*{1cm}\times(-1)^{M-K}\underbrace{\left(
\begin{array}{ccc}
J   &   2   &   J^{\prime}\\
-M  &   0   &   M
\end{array}\right)}_{J^{\prime}=J,J\pm1,J\pm2}
\underbrace{\left(
\begin{array}{ccc}
J   &   2   &   J^{\prime}\\
-K  &   0   &   K
\end{array}\right)}_{J^{\prime}=J,J\pm,J\pm2}
\end{align*}

If, on the other hand, the laser field is tilted relative to the fixed static field, the $\cos^2\theta_L$ operator becomes
\begin{align*}
\cos^2\theta_L=&\cos^2\beta\cos^2\theta+2\sin\beta\cos\beta\sin\theta\cos\theta\cos\varphi\nonumber\\&+\sin^2\beta\sin^2\theta\cos^2\varphi\\
=&\frac{1}{3}\!+\! \frac{2}{3}\cos^2\beta D^2_{00}\!+\!2\sqrt{\frac{1}{6}}\sin\beta\cos\beta\left(D^2_{-\!10}\!-\!D^2_{10}\right)\nonumber\\ &-\frac{1}{3}D^2_{00} \sin^2\beta+\sqrt{\frac{1}{6}}\sin^2\beta\left(D^2_{-20}+D^2_{20}\right)
\end{align*}
The selection rules then become: $J^{\prime}=J,J\pm1,J\pm2$ and $M^{\prime}=M,M\pm1,M\pm2$. The equivalence of the two choices
of expressing the fields has been numerically checked, but only the first, `simple' choice has been used in the calculations.

\end{appendix}

\bibliography{HaerFri08}

\begin{thebibliography}{40}
\expandafter\ifx\csname natexlab\endcsname\relax\def\natexlab#1{#1}\fi
\expandafter\ifx\csname bibnamefont\endcsname\relax
  \def\bibnamefont#1{#1}\fi
\expandafter\ifx\csname bibfnamefont\endcsname\relax
  \def\bibfnamefont#1{#1}\fi
\expandafter\ifx\csname citenamefont\endcsname\relax
  \def\citenamefont#1{#1}\fi
\expandafter\ifx\csname url\endcsname\relax
  \def\url#1{\texttt{#1}}\fi
\expandafter\ifx\csname urlprefix\endcsname\relax\def\urlprefix{URL }\fi
\providecommand{\bibinfo}[2]{#2}
\providecommand{\eprint}[2][]{\url{#2}}

\bibitem[{\citenamefont{Herschbach}(2006)}]{special}
\bibinfo{author}{\bibfnamefont{D.}~\bibnamefont{Herschbach}},
  \bibinfo{journal}{European Physical Journal D} \textbf{\bibinfo{volume}{38}},
  \bibinfo{pages}{3} (\bibinfo{year}{2006}).

\bibitem[{\citenamefont{Friedrich and Herschbach}(1991)}]{Friedrich1991}
\bibinfo{author}{\bibfnamefont{B.}~\bibnamefont{Friedrich}} \bibnamefont{and}
  \bibinfo{author}{\bibfnamefont{D.~R.} \bibnamefont{Herschbach}},
  \bibinfo{journal}{Nature} \textbf{\bibinfo{volume}{353}},
  \bibinfo{pages}{412} (\bibinfo{year}{1991}).

\bibitem[{\citenamefont{Loesch and Remscheid}(1990)}]{LOESCH1990}
\bibinfo{author}{\bibfnamefont{H.~J.} \bibnamefont{Loesch}} \bibnamefont{and}
  \bibinfo{author}{\bibfnamefont{A.}~\bibnamefont{Remscheid}},
  \bibinfo{journal}{J. Chem. Phys.} \textbf{\bibinfo{volume}{93}},
  \bibinfo{pages}{4779} (\bibinfo{year}{1990}).

\bibitem[{\citenamefont{Friedrich and
  Herschbach}(1995{\natexlab{a}})}]{FRIEDRICH1995}
\bibinfo{author}{\bibfnamefont{B.}~\bibnamefont{Friedrich}} \bibnamefont{and}
  \bibinfo{author}{\bibfnamefont{D.}~\bibnamefont{Herschbach}},
  \bibinfo{journal}{Phys. Rev. Lett.} \textbf{\bibinfo{volume}{74}},
  \bibinfo{pages}{4623} (\bibinfo{year}{1995}{\natexlab{a}}).

\bibitem[{\citenamefont{Friedrich and
  Herschbach}(1995{\natexlab{b}})}]{FRIEDRICH1995a}
\bibinfo{author}{\bibfnamefont{B.}~\bibnamefont{Friedrich}} \bibnamefont{and}
  \bibinfo{author}{\bibfnamefont{D.}~\bibnamefont{Herschbach}},
  \bibinfo{journal}{J. Phys. Chem.} \textbf{\bibinfo{volume}{99}},
  \bibinfo{pages}{15686} (\bibinfo{year}{1995}{\natexlab{b}}).

\bibitem[{\citenamefont{Stapelfeldt and Seideman}(2003)}]{Colloquium2003}
\bibinfo{author}{\bibfnamefont{H.}~\bibnamefont{Stapelfeldt}} \bibnamefont{and}
  \bibinfo{author}{\bibfnamefont{T.}~\bibnamefont{Seideman}},
  \bibinfo{journal}{Rev.Mod.Phys.} \textbf{\bibinfo{volume}{75}},
  \bibinfo{pages}{543} (\bibinfo{year}{2003}), \bibinfo{note}{and refs. cited
  therein}.

\bibitem[{\citenamefont{Bandrauk et~al.}(2002)\citenamefont{Bandrauk, Fujimura,
  and Gordon}}]{LasConManMol}
\bibinfo{author}{\bibfnamefont{A.~D.} \bibnamefont{Bandrauk}},
  \bibinfo{author}{\bibfnamefont{Y.}~\bibnamefont{Fujimura}}, \bibnamefont{and}
  \bibinfo{author}{\bibfnamefont{R.~J.} \bibnamefont{Gordon}},
  \emph{\bibinfo{title}{{Laser Control and Manipulation of Molecules \#821}}}
  (\bibinfo{publisher}{An American Chemical Society Publication},
  \bibinfo{year}{2002}), ISBN \bibinfo{isbn}{0841237867}, \bibinfo{note}{and
  refs. cited therein}.

\bibitem[{\citenamefont{Friedrich and Herschbach}(1996)}]{Friedrich1996}
\bibinfo{author}{\bibfnamefont{B.}~\bibnamefont{Friedrich}} \bibnamefont{and}
  \bibinfo{author}{\bibfnamefont{D.}~\bibnamefont{Herschbach}},
  \bibinfo{journal}{Chem. Phys. Lett.} \textbf{\bibinfo{volume}{262}},
  \bibinfo{pages}{41} (\bibinfo{year}{1996}).

\bibitem[{\citenamefont{Kim and Felker}(1997)}]{Kim1997}
\bibinfo{author}{\bibfnamefont{W.~S.} \bibnamefont{Kim}} \bibnamefont{and}
  \bibinfo{author}{\bibfnamefont{P.~M.} \bibnamefont{Felker}},
  \bibinfo{journal}{J. Chem. Phys.} \textbf{\bibinfo{volume}{107}},
  \bibinfo{pages}{2193} (\bibinfo{year}{1997}).

\bibitem[{\citenamefont{Larsen et~al.}(2000)\citenamefont{Larsen, Hald, Bjerre,
  Stapelfeldt, and Seideman}}]{Larsen2000}
\bibinfo{author}{\bibfnamefont{J.~J.} \bibnamefont{Larsen}},
  \bibinfo{author}{\bibfnamefont{K.}~\bibnamefont{Hald}},
  \bibinfo{author}{\bibfnamefont{N.}~\bibnamefont{Bjerre}},
  \bibinfo{author}{\bibfnamefont{H.}~\bibnamefont{Stapelfeldt}},
  \bibnamefont{and} \bibinfo{author}{\bibfnamefont{T.}~\bibnamefont{Seideman}},
  \bibinfo{journal}{Phys. Rev. Lett.} \textbf{\bibinfo{volume}{85}},
  \bibinfo{pages}{2470} (\bibinfo{year}{2000}).

\bibitem[{\citenamefont{Poulsen et~al.}(2004)\citenamefont{Poulsen, Peronne,
  Stapelfeldt, Bisgaard, Viftrup, Hamilton, and Seideman}}]{Poulsen2004}
\bibinfo{author}{\bibfnamefont{M.~D.} \bibnamefont{Poulsen}},
  \bibinfo{author}{\bibfnamefont{E.}~\bibnamefont{Peronne}},
  \bibinfo{author}{\bibfnamefont{H.}~\bibnamefont{Stapelfeldt}},
  \bibinfo{author}{\bibfnamefont{C.~Z.} \bibnamefont{Bisgaard}},
  \bibinfo{author}{\bibfnamefont{S.~S.} \bibnamefont{Viftrup}},
  \bibinfo{author}{\bibfnamefont{E.}~\bibnamefont{Hamilton}}, \bibnamefont{and}
  \bibinfo{author}{\bibfnamefont{T.}~\bibnamefont{Seideman}},
  \bibinfo{journal}{J. Chem. Phys.} \textbf{\bibinfo{volume}{121}},
  \bibinfo{pages}{783} (\bibinfo{year}{2004}).

\bibitem[{\citenamefont{Zhao et~al.}(2003)\citenamefont{Zhao, Lee, Chung, S.,
  Kang, Friedrich, and Chung}}]{Zhao2003}
\bibinfo{author}{\bibfnamefont{B.~S.} \bibnamefont{Zhao}},
  \bibinfo{author}{\bibfnamefont{S.~H.} \bibnamefont{Lee}},
  \bibinfo{author}{\bibfnamefont{H.~S.} \bibnamefont{Chung}},
  \bibinfo{author}{\bibfnamefont{H.}~\bibnamefont{S.}},
  \bibinfo{author}{\bibfnamefont{W.~K.} \bibnamefont{Kang}},
  \bibinfo{author}{\bibfnamefont{B.}~\bibnamefont{Friedrich}},
  \bibnamefont{and} \bibinfo{author}{\bibfnamefont{D.~S.} \bibnamefont{Chung}},
  \bibinfo{journal}{J. Chem. Phys.} \textbf{\bibinfo{volume}{119}},
  \bibinfo{pages}{8905} (\bibinfo{year}{2003}).

\bibitem[{\citenamefont{Fulton et~al.}(2006)\citenamefont{Fulton, Bishop,
  Shneider, and Barker}}]{Fulton2006}
\bibinfo{author}{\bibfnamefont{R.}~\bibnamefont{Fulton}},
  \bibinfo{author}{\bibfnamefont{A.~I.} \bibnamefont{Bishop}},
  \bibinfo{author}{\bibfnamefont{M.~N.} \bibnamefont{Shneider}},
  \bibnamefont{and} \bibinfo{author}{\bibfnamefont{P.~F.}
  \bibnamefont{Barker}}, \bibinfo{journal}{Nature Physics}
  \textbf{\bibinfo{volume}{2}}, \bibinfo{pages}{465} (\bibinfo{year}{2006}).

\bibitem[{\citenamefont{Baumfalk et~al.}(2001)\citenamefont{Baumfalk, Nahler,
  and Buck}}]{Baumfalk2001}
\bibinfo{author}{\bibfnamefont{R.}~\bibnamefont{Baumfalk}},
  \bibinfo{author}{\bibfnamefont{N.~H.} \bibnamefont{Nahler}},
  \bibnamefont{and} \bibinfo{author}{\bibfnamefont{U.}~\bibnamefont{Buck}},
  \bibinfo{journal}{J. Chem. Phys.} \textbf{\bibinfo{volume}{114}},
  \bibinfo{pages}{4755} (\bibinfo{year}{2001}).

\bibitem[{\citenamefont{Nahler et~al.}(2003)\citenamefont{Nahler, Baumfalk, and
  Buck}}]{Buck03-01}
\bibinfo{author}{\bibfnamefont{H.}~\bibnamefont{Nahler}},
  \bibinfo{author}{\bibfnamefont{R.}~\bibnamefont{Baumfalk}}, \bibnamefont{and}
  \bibinfo{author}{\bibfnamefont{U.}~\bibnamefont{Buck}},
  \bibinfo{journal}{J.Chem.Phys.} \textbf{\bibinfo{volume}{119}},
  \bibinfo{pages}{224} (\bibinfo{year}{2003}).

\bibitem[{\citenamefont{Friedrich et~al.}(2003)\citenamefont{Friedrich, Nahler,
  and Buck}}]{Friedrich2003}
\bibinfo{author}{\bibfnamefont{B.}~\bibnamefont{Friedrich}},
  \bibinfo{author}{\bibfnamefont{N.~H.} \bibnamefont{Nahler}},
  \bibnamefont{and} \bibinfo{author}{\bibfnamefont{U.}~\bibnamefont{Buck}},
  \bibinfo{journal}{J. Mod. Opt.} \textbf{\bibinfo{volume}{50}},
  \bibinfo{pages}{2677} (\bibinfo{year}{2003}).

\bibitem[{\citenamefont{Poterya et~al.}(2008)\citenamefont{Poterya, Votava,
  Farnik, Oncak, Slavicek, Buck, and Friedrich}}]{farnik}
\bibinfo{author}{\bibfnamefont{V.}~\bibnamefont{Poterya}},
  \bibinfo{author}{\bibfnamefont{O.}~\bibnamefont{Votava}},
  \bibinfo{author}{\bibfnamefont{M.}~\bibnamefont{Farnik}},
  \bibinfo{author}{\bibfnamefont{M.}~\bibnamefont{Oncak}},
  \bibinfo{author}{\bibfnamefont{P.}~\bibnamefont{Slavicek}},
  \bibinfo{author}{\bibfnamefont{U.}~\bibnamefont{Buck}}, \bibnamefont{and}
  \bibinfo{author}{\bibfnamefont{B.}~\bibnamefont{Friedrich}},
  \bibinfo{journal}{J. Chem. Phys.} \textbf{\bibinfo{volume}{128}},
  \bibinfo{pages}{104313} (\bibinfo{year}{2008}).

\bibitem[{\citenamefont{Sakai et~al.}(2003)\citenamefont{Sakai, Minemoto,
  Nanjo, Tanji, and Suzuki}}]{Sakai2003}
\bibinfo{author}{\bibfnamefont{H.}~\bibnamefont{Sakai}},
  \bibinfo{author}{\bibfnamefont{S.}~\bibnamefont{Minemoto}},
  \bibinfo{author}{\bibfnamefont{H.}~\bibnamefont{Nanjo}},
  \bibinfo{author}{\bibfnamefont{H.}~\bibnamefont{Tanji}}, \bibnamefont{and}
  \bibinfo{author}{\bibfnamefont{T.}~\bibnamefont{Suzuki}},
  \bibinfo{journal}{Phys. Rev. Lett.} \textbf{\bibinfo{volume}{90}},
  \bibinfo{pages}{083001} (\bibinfo{year}{2003}).

\bibitem[{\citenamefont{Tanji et~al.}(2005)\citenamefont{Tanji, Minemoto, and
  Sakai}}]{Tanji2005}
\bibinfo{author}{\bibfnamefont{H.}~\bibnamefont{Tanji}},
  \bibinfo{author}{\bibfnamefont{S.}~\bibnamefont{Minemoto}}, \bibnamefont{and}
  \bibinfo{author}{\bibfnamefont{H.}~\bibnamefont{Sakai}},
  \bibinfo{journal}{Physical Review A} \textbf{\bibinfo{volume}{72}},
  \bibinfo{pages}{063401} (\bibinfo{year}{2005}).

\bibitem[{\citenamefont{Friedrich and
  Herschbach}(1999{\natexlab{a}})}]{FriedHerschb99-1}
\bibinfo{author}{\bibfnamefont{B.}~\bibnamefont{Friedrich}} \bibnamefont{and}
  \bibinfo{author}{\bibfnamefont{D.}~\bibnamefont{Herschbach}},
  \bibinfo{journal}{J.Phys.Chem.A} \textbf{\bibinfo{volume}{103}},
  \bibinfo{pages}{10280} (\bibinfo{year}{1999}{\natexlab{a}}).

\bibitem[{\citenamefont{Friedrich and
  Herschbach}(1999{\natexlab{b}})}]{FriedHerschb99-2}
\bibinfo{author}{\bibfnamefont{B.}~\bibnamefont{Friedrich}} \bibnamefont{and}
  \bibinfo{author}{\bibfnamefont{D.}~\bibnamefont{Herschbach}},
  \bibinfo{journal}{J. Chem. Phys.} \textbf{\bibinfo{volume}{111}},
  \bibinfo{pages}{6157} (\bibinfo{year}{1999}{\natexlab{b}}).

\bibitem[{\citenamefont{Kim and Felker}(1998)}]{KimFelker98-1}
\bibinfo{author}{\bibfnamefont{W.}~\bibnamefont{Kim}} \bibnamefont{and}
  \bibinfo{author}{\bibfnamefont{P.}~\bibnamefont{Felker}},
  \bibinfo{journal}{J. Chem. Phys.} \textbf{\bibinfo{volume}{108}},
  \bibinfo{pages}{6763} (\bibinfo{year}{1998}).

\bibitem[{\citenamefont{Dion et~al.}(1999)\citenamefont{Dion, Keller, Atabek,
  and Bandrauk}}]{Dion1999}
\bibinfo{author}{\bibfnamefont{C.~M.} \bibnamefont{Dion}},
  \bibinfo{author}{\bibfnamefont{A.}~\bibnamefont{Keller}},
  \bibinfo{author}{\bibfnamefont{O.}~\bibnamefont{Atabek}}, \bibnamefont{and}
  \bibinfo{author}{\bibfnamefont{A.~D.} \bibnamefont{Bandrauk}},
  \bibinfo{journal}{Phys. Rev. A} \textbf{\bibinfo{volume}{59}},
  \bibinfo{pages}{1382} (\bibinfo{year}{1999}).

\bibitem[{\citenamefont{Henriksen}(1999)}]{Henriksen1999}
\bibinfo{author}{\bibfnamefont{N.~E.} \bibnamefont{Henriksen}},
  \bibinfo{journal}{Chem. Phys. Lett.} \textbf{\bibinfo{volume}{312}},
  \bibinfo{pages}{196} (\bibinfo{year}{1999}).

\bibitem[{\citenamefont{Bunker and Jensen}(1998)}]{BunkerJensen98}
\bibinfo{author}{\bibfnamefont{P.}~\bibnamefont{Bunker}} \bibnamefont{and}
  \bibinfo{author}{\bibfnamefont{P.}~\bibnamefont{Jensen}},
  \emph{\bibinfo{title}{{Molecular Symmetry \& Spectroscopy}}}
  (\bibinfo{publisher}{NRC Press (Canada)}, \bibinfo{year}{1998}), ISBN
  \bibinfo{isbn}{0660175193}.

\bibitem[{\citenamefont{Tinkham}(1964)}]{Tinkham}
\bibinfo{author}{\bibfnamefont{M.}~\bibnamefont{Tinkham}},
  \emph{\bibinfo{title}{{Group Theory and Quantum Mechanics}}}
  (\bibinfo{publisher}{McGraw-Hill Book Company}, \bibinfo{year}{1964}).

\bibitem[{\citenamefont{Biedenharn and Rose}(1953)}]{BiedenharnRose}
\bibinfo{author}{\bibfnamefont{L.}~\bibnamefont{Biedenharn}} \bibnamefont{and}
  \bibinfo{author}{\bibfnamefont{M.}~\bibnamefont{Rose}},
  \bibinfo{journal}{Rev.Mod.Phys.} \textbf{\bibinfo{volume}{25}},
  \bibinfo{pages}{729} (\bibinfo{year}{1953}).

\bibitem[{\citenamefont{Rost et~al.}(1992)\citenamefont{Rost, Griffin,
  Friedrich, and Herschbach}}]{ROST1992}
\bibinfo{author}{\bibfnamefont{J.~M.} \bibnamefont{Rost}},
  \bibinfo{author}{\bibfnamefont{J.~C.} \bibnamefont{Griffin}},
  \bibinfo{author}{\bibfnamefont{B.}~\bibnamefont{Friedrich}},
  \bibnamefont{and} \bibinfo{author}{\bibfnamefont{D.~R.}
  \bibnamefont{Herschbach}}, \bibinfo{journal}{Phys. Rev. Lett.}
  \textbf{\bibinfo{volume}{68}}, \bibinfo{pages}{1299} (\bibinfo{year}{1992}).

\bibitem[{\citenamefont{Maergoiz and Troe}(1993)}]{MAERGOIZ1993}
\bibinfo{author}{\bibfnamefont{A.~I.} \bibnamefont{Maergoiz}} \bibnamefont{and}
  \bibinfo{author}{\bibfnamefont{J.}~\bibnamefont{Troe}}, \bibinfo{journal}{J.
  Chem. Phys.} \textbf{\bibinfo{volume}{99}}, \bibinfo{pages}{3218}
  (\bibinfo{year}{1993}).

\bibitem[{\citenamefont{Friedrich et~al.}(1994)\citenamefont{Friedrich,
  Slenczka, and Herschbach}}]{frislenher}
\bibinfo{author}{\bibfnamefont{B.}~\bibnamefont{Friedrich}},
  \bibinfo{author}{\bibfnamefont{A.}~\bibnamefont{Slenczka}}, \bibnamefont{and}
  \bibinfo{author}{\bibfnamefont{D.}~\bibnamefont{Herschbach}},
  \bibinfo{journal}{Can. J. Phys.} \textbf{\bibinfo{volume}{72}},
  \bibinfo{pages}{897} (\bibinfo{year}{1994}).

\bibitem[{\citenamefont{Arango et~al.}(2004)\citenamefont{Arango, Kennerly, and
  Ezra}}]{Arango2004}
\bibinfo{author}{\bibfnamefont{C.~A.} \bibnamefont{Arango}},
  \bibinfo{author}{\bibfnamefont{W.~W.} \bibnamefont{Kennerly}},
  \bibnamefont{and} \bibinfo{author}{\bibfnamefont{G.~S.} \bibnamefont{Ezra}},
  \bibinfo{journal}{Chem. Phys. Lett.} \textbf{\bibinfo{volume}{392}},
  \bibinfo{pages}{486} (\bibinfo{year}{2004}).

\bibitem[{\citenamefont{Kozin and Roberts}(2003)}]{Kozin2003}
\bibinfo{author}{\bibfnamefont{I.~N.} \bibnamefont{Kozin}} \bibnamefont{and}
  \bibinfo{author}{\bibfnamefont{R.~M.} \bibnamefont{Roberts}},
  \bibinfo{journal}{J. Chem. Phys.} \textbf{\bibinfo{volume}{118}},
  \bibinfo{pages}{10523} (\bibinfo{year}{2003}).

\bibitem[{\citenamefont{Sadovskii and Zhilinskii}(2006)}]{Sadovskii2006}
\bibinfo{author}{\bibfnamefont{D.~A.} \bibnamefont{Sadovskii}}
  \bibnamefont{and} \bibinfo{author}{\bibfnamefont{B.~I.}
  \bibnamefont{Zhilinskii}}, \bibinfo{journal}{Mol. Phys.}
  \textbf{\bibinfo{volume}{104}}, \bibinfo{pages}{2595} (\bibinfo{year}{2006}).

\bibitem[{\citenamefont{Arango et~al.}(2005)\citenamefont{Arango, Kennerly, and
  Ezra}}]{Arango05}
\bibinfo{author}{\bibfnamefont{C.~A.} \bibnamefont{Arango}},
  \bibinfo{author}{\bibfnamefont{W.~W.} \bibnamefont{Kennerly}},
  \bibnamefont{and} \bibinfo{author}{\bibfnamefont{G.~S.} \bibnamefont{Ezra}},
  \bibinfo{journal}{J. Chem. Phys.} \textbf{\bibinfo{volume}{122}},
  \bibinfo{pages}{184303} (\bibinfo{year}{2005}).

\bibitem[{\citenamefont{Kennerly}(2005)}]{PhdKennerly05}
\bibinfo{author}{\bibfnamefont{W.}~\bibnamefont{Kennerly}}, Ph.D. thesis,
  \bibinfo{school}{Faculty of the Graduate School of Cornell University}
  (\bibinfo{year}{2005}).

\bibitem[{\citenamefont{Pogliani}(2003)}]{Pogliani2003}
\bibinfo{author}{\bibfnamefont{L.}~\bibnamefont{Pogliani}},
  \bibinfo{journal}{New J. Chem.} \textbf{\bibinfo{volume}{27}},
  \bibinfo{pages}{919} (\bibinfo{year}{2003}).

\bibitem[{Bei()}]{Beilstein}
\bibinfo{howpublished}{CrossFire Beilstein database}.

\bibitem[{\citenamefont{Schulz et~al.}(2004)\citenamefont{Schulz, Bethlem, van
  Veldhoven, Kupper, Conrad, and Meijer}}]{Schulz2004}
\bibinfo{author}{\bibfnamefont{S.~A.} \bibnamefont{Schulz}},
  \bibinfo{author}{\bibfnamefont{H.~L.} \bibnamefont{Bethlem}},
  \bibinfo{author}{\bibfnamefont{J.}~\bibnamefont{van Veldhoven}},
  \bibinfo{author}{\bibfnamefont{J.}~\bibnamefont{Kupper}},
  \bibinfo{author}{\bibfnamefont{H.}~\bibnamefont{Conrad}}, \bibnamefont{and}
  \bibinfo{author}{\bibfnamefont{G.}~\bibnamefont{Meijer}},
  \bibinfo{journal}{Phys. Rev. Lett.} \textbf{\bibinfo{volume}{93}},
  \bibinfo{pages}{020406} (\bibinfo{year}{2004}).

\bibitem[{\citenamefont{Buck and Farnik}(2006)}]{Buck2006}
\bibinfo{author}{\bibfnamefont{U.}~\bibnamefont{Buck}} \bibnamefont{and}
  \bibinfo{author}{\bibfnamefont{M.}~\bibnamefont{Farnik}},
  \bibinfo{journal}{Int. Rev. Phys. Chem.} \textbf{\bibinfo{volume}{25}},
  \bibinfo{pages}{583} (\bibinfo{year}{2006}).

\bibitem[{\citenamefont{Wu et~al.}(1994)\citenamefont{Wu, Bemish, and
  Miller}}]{Wu1994}
\bibinfo{author}{\bibfnamefont{M.}~\bibnamefont{Wu}},
  \bibinfo{author}{\bibfnamefont{R.~J.} \bibnamefont{Bemish}},
  \bibnamefont{and} \bibinfo{author}{\bibfnamefont{R.~E.}
  \bibnamefont{Miller}}, \bibinfo{journal}{J. Chem. Phys.}
  \textbf{\bibinfo{volume}{101}}, \bibinfo{pages}{9447} (\bibinfo{year}{1994}).

\end{thebibliography}

\end{document}